\documentclass[twocolumn]{aastex631}

\usepackage{booktabs}
\usepackage{comment}
\usepackage{enumitem}   
\begin{document}

\title{Hybrid Morphology Radio Sources from the MeerKAT Absorption Line Survey (MALS): Radio, Mid-infrared, and Environmental Characteristics}

\shorttitle{HyMoRS from MALS}
\shortauthors{Manik et al.}

\author[0000-0002-6794-7405]{Souvik Manik}
\affiliation{Midnapore City College, Kuturia, Bhadutala, Paschim Medinipur, West Bengal, 721129, India}
\email{Email:souvikmanik12@gmail.com}

\author[0000-0003-4213-9679]{Shobha Kumari}
\affiliation{Midnapore City College, Kuturia, Bhadutala, Paschim Medinipur, West Bengal, 721129, India}

\author[0000-0002-1829-847X]{Netai Bhukta}
\affiliation{Department of Physics, Sidho Kanho Birsha University, Ranchi Road, Purulia, 723104, India}

\author[0000-0003-2325-8509]{Sabyasachi Pal}
\affiliation{Midnapore City College, Kuturia, Bhadutala, Paschim Medinipur, West Bengal, 721129, India}

\author[0000-0003-4353-8487]{Sushanta K. Mondal}
\affiliation{Department of Physics, Sidho Kanho Birsha University, Ranchi Road, Purulia, 723104, India}

\begin{abstract}
Hybrid morphology radio sources (HyMoRSs) are a rare subclass of radio galaxies that display a Fanaroff-Riley type I (FR I) morphology on one side of the central supermassive black hole (SMBH) and a type II (FR II) morphology on the other. In this study, we report the discovery of thirty-six new HyMoRSs, marking the largest collection of such sources in the southern sky to date, using data obtained from the MeerKAT absorption line survey (MALS). The identified HyMoRSs exhibit moderate radio luminosities in the range \(9.9 \times 10^{23}\) to \(5.7 \times 10^{25} \, \mathrm{W\,Hz^{-1}}\), with a median value of \(4.4 \times 10^{24} \, \mathrm{W\,Hz^{-1}}\) at 1.4 GHz, and are located within the redshift range \(0.04<z<1.34\). In this work, we show for the first time that the two lobes of HyMoRSs exhibit no statistically significant difference in their spectral indices. We also investigate the mid-infrared properties and environments of their host galaxies. Notably, nine out of the thirty-six sources are situated near the centers of galaxy clusters, including one with giant radio jets that extend over 811 kpc. Our analysis reveals that the majority of HyMoRSs are hosted by actively star-forming galaxies that exhibit elevated star formation rates. Furthermore, our findings suggest that HyMoRSs may arise from FR II jets being deflected by a dense, cluster-like environment, along with orientation effects that make one jet appear FR I-like. As our candidates are selected through visual inspection of MALS radio maps, higher-resolution follow-up observations are still necessary to confirm the nature of their morphologies.
\end{abstract}


\section{Introduction} 
\label{sec:intro}
Radio galaxies (RG) contain active galactic nuclei (AGN) that usually produce significant radio emissions in the form of twin symmetrical relativistic jets that extend far beyond their host galaxy over several hundred kiloparsec to megaparsec. In general, morphologically RG was classified into Fanaroff–Riley class I (FR I) and class II (FR II) \citep{Fa74}.
The FR ratio (the distance between the regions of peak brightness on opposite sides of the core galaxy to the total extent of the source measured from the lowest contour) was used to classify radio galaxies \citep{Fa74}. Radio galaxies with FR ratios less than 0.5 were categorised as FR I, whereas those with FR ratios larger than 0.5 were classified as FR II. With high-resolution images, radio galaxies with hot spots (the brightest regions) close to their center galaxies are classified as FR I and those with hotspots at the outer edge are considered FR II \citep{Ma71}. The radio galaxies are also categorized based on radio power. $L_{\textrm{178 MHz}}<2\times10^{25}$ W Hz$^{-1}$sr$^{-1}$ are FR I, while those with higher radio power are FR II \citep{Fa74}. 

The FR I/FR II dichotomy classification of extra-galactic radio sources is a contentious topic in extra-galactic astronomy. It is believed that the morphological differences in FR I and FR II sources are related to the (i) transition of an initially supersonic (relatively weak) jet to a transonic/subsonic flow which is substantially decelerated by thermal plasma within the inner region of the central host galaxy \citep{Ko88, Bo96, Bi95, Ka97}; (ii) the nature of the central black hole and the composition of jets, i.e. e$^{-}$ -- e$^{+}$ plasma for FR I sources, whereas e$^{-}$ -- p could be preferred for the case of FR II sources \citep{Re96a, Re96b, Me97, Ce97, Me99}; and (iii) the density jump towards the asymmetric environment of the external medium. Hybrid morphology radio sources (HyMoRS), which exhibit FR~I structure on one side of the radio core and FR~II structure on the other, offer a unique opportunity to investigate the physical conditions underlying the FR dichotomy. The term ``HyMoRS'' was first introduced by \citet{Go00}, following the identification of six such sources using the Giant Metrewave Radio Telescope (GMRT). \citet{Ga06, Ka17, Pa22} identified a total of sixty-three HyMoRS using the Faint Images of the Radio Sky at Twenty-cm (FIRST) survey \citep{Be95, Wh97}. The number of HyMoRS candidates has rapidly grown over the past few years, amounting to nearly sixty sources\citep{Mi19, Ka17, Pa22}.

\begin{figure*}
\vbox{
\centerline{
\includegraphics[width=9.5cm, origin=c]{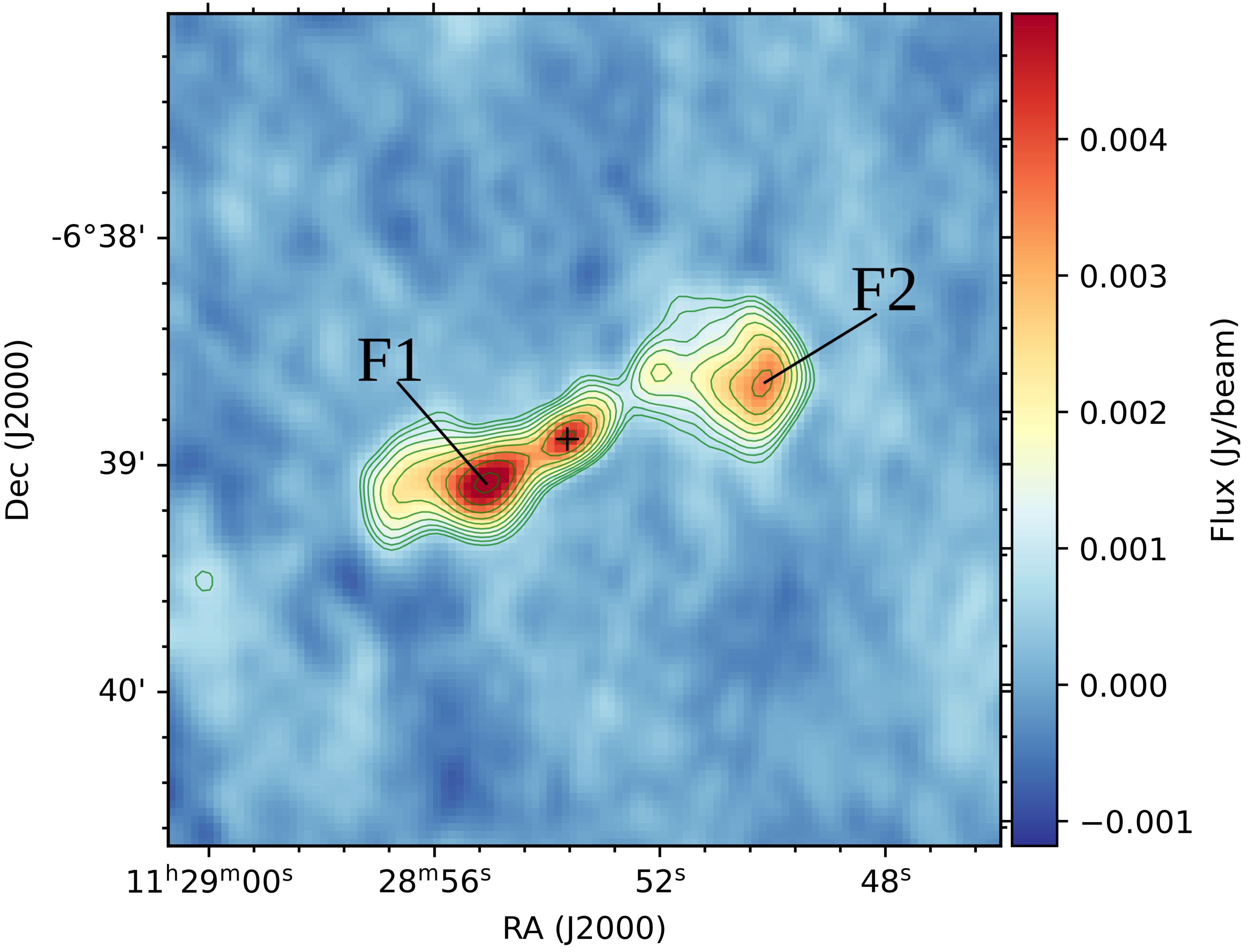}
\includegraphics[width=9.5cm, height=7cm, origin=c]{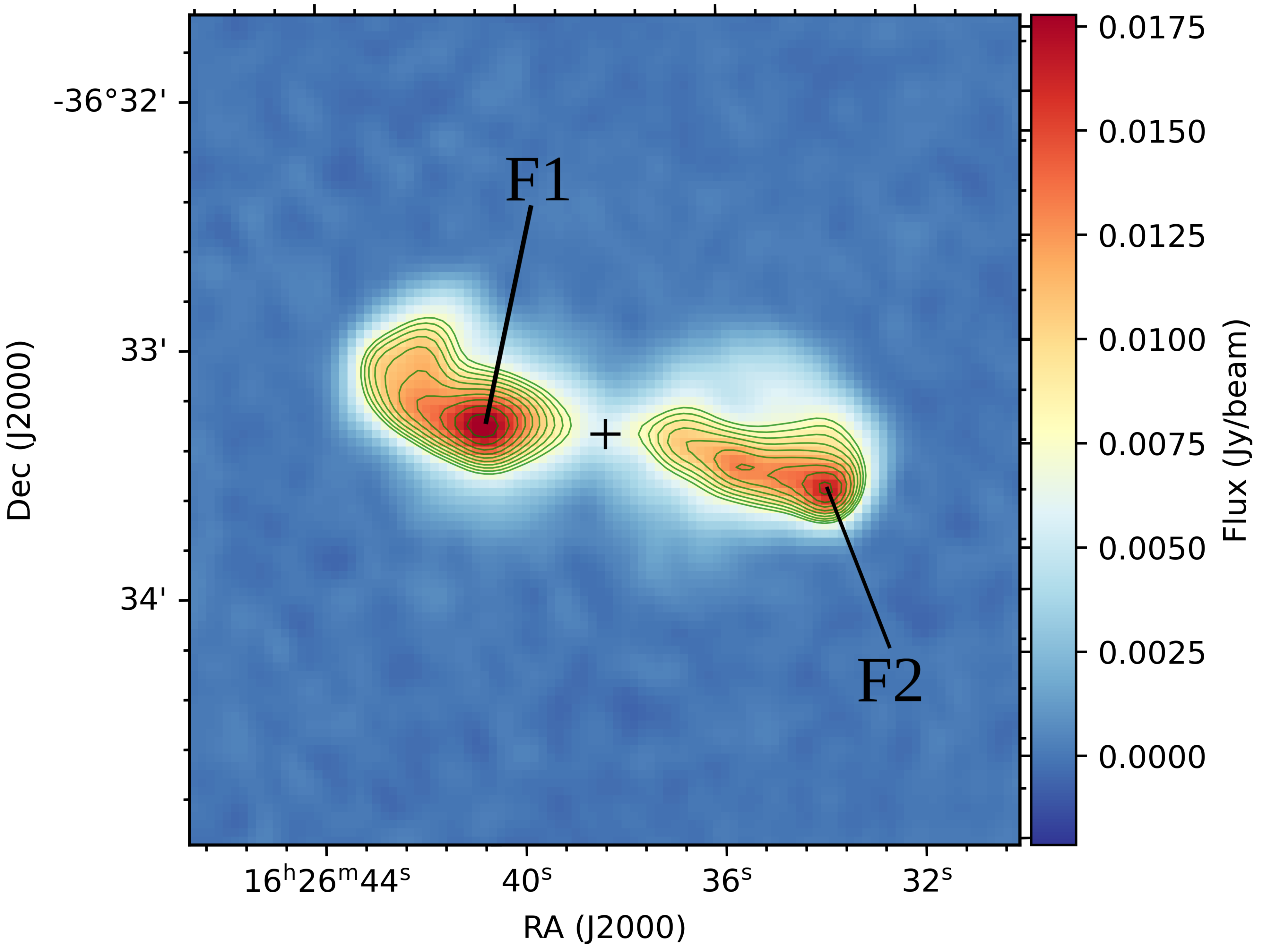}
}
}
\vbox{
\centerline{
\includegraphics[width=9.5cm, height=6.75cm, origin=c]{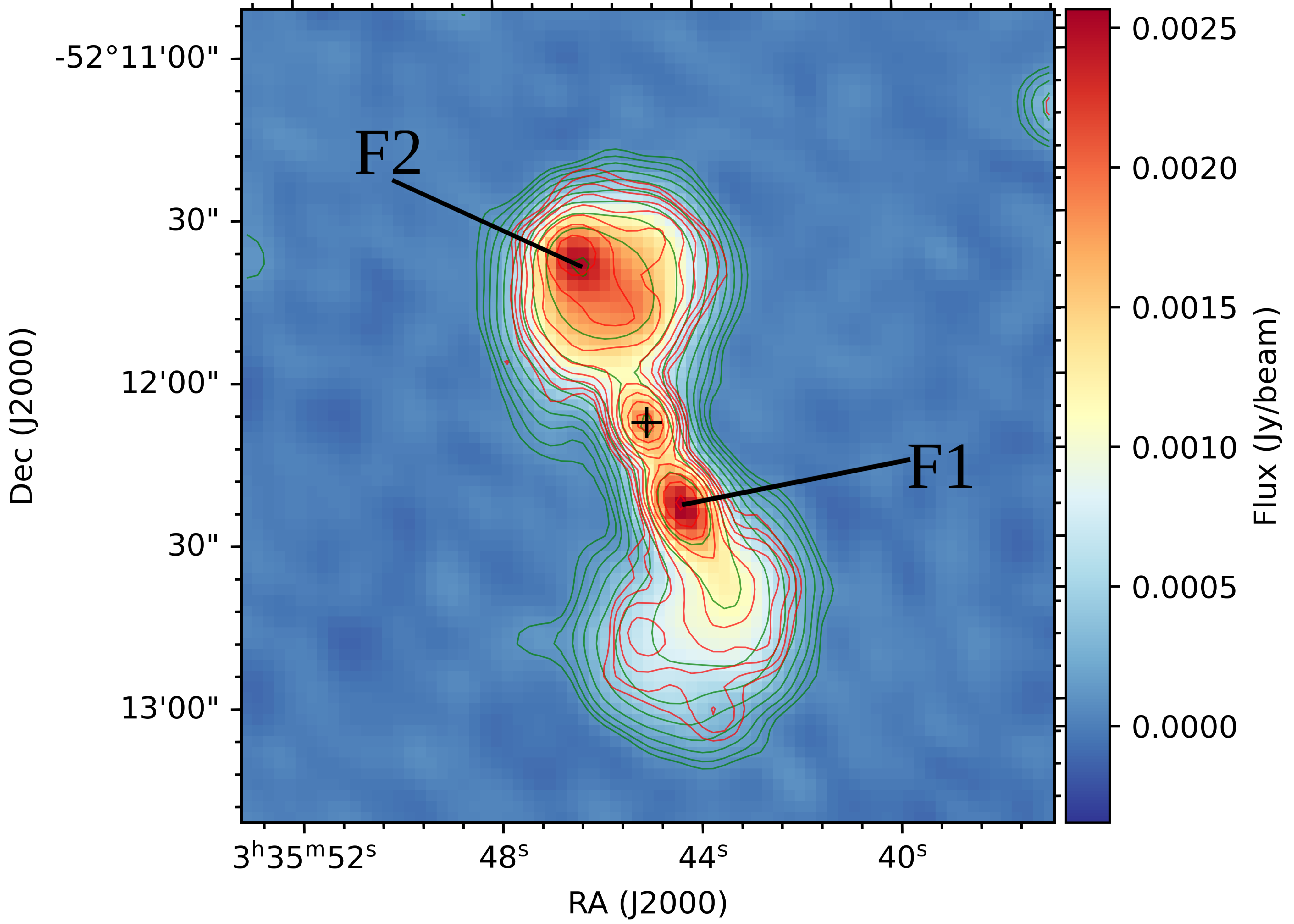}
\includegraphics[width=9.5cm, origin=c]{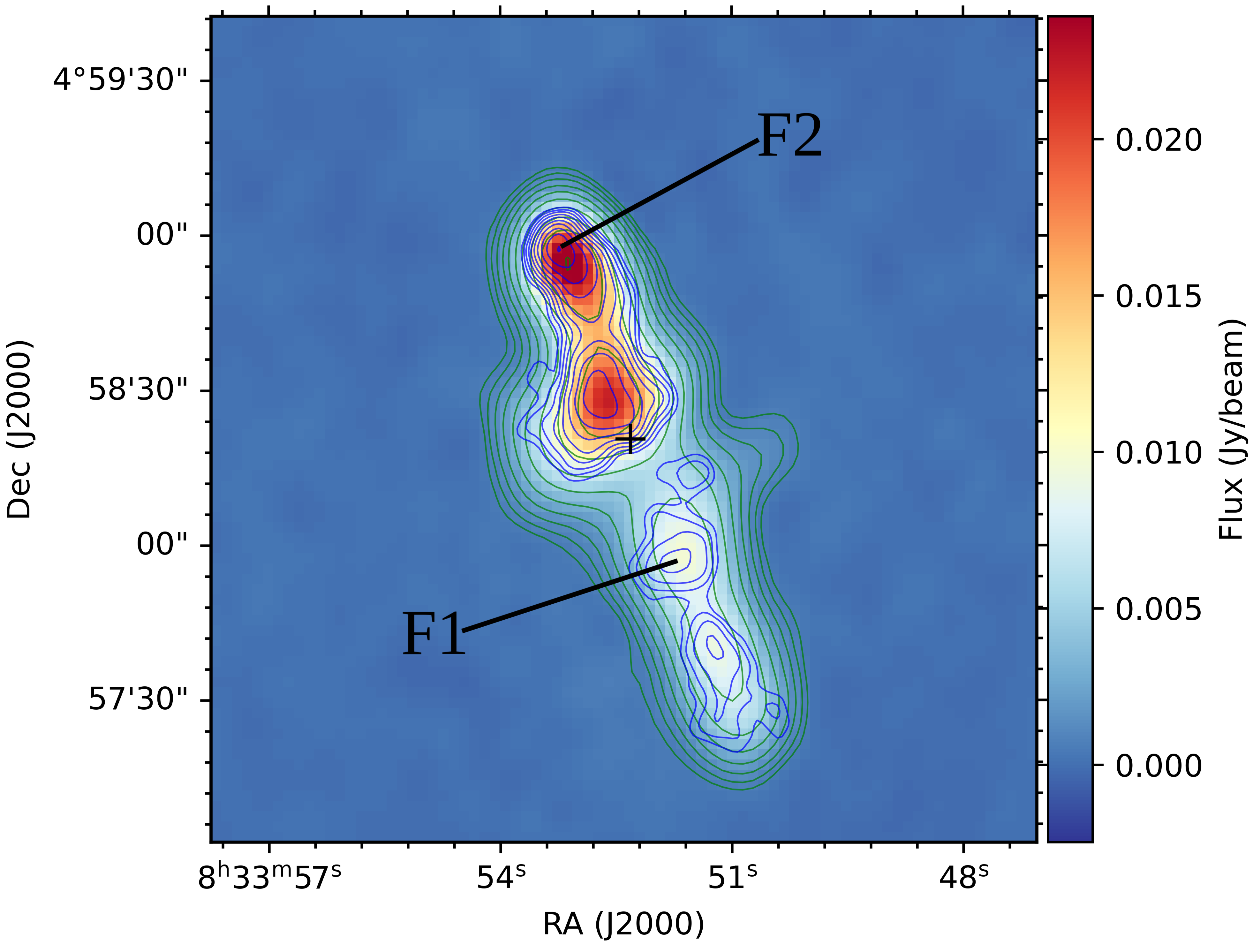}
}
}
\caption{A sample of four HyMoRSs identified from MALS at 1000 MHz and 1400 MHz. Sources are superimposed with 1006 MHz and 1381 MHz of the MAL survey (in green and red contours), and 1400 MHz of the FIRST survey (blue contours).}
\label{fig:sample}
\end{figure*}

This paper is organized as follows: In Section \ref{sec:style}, we explain the search procedure for HyMoRS. In Section \ref{sec:result}, we explain our results and discuss them in Section \ref{sec:discuss}. Section \ref{sec:conclusion} summarizes this paper.
We used the following cosmology parameters in the current paper from the result of final full-mission Planck measurements of the CMB anisotropies: $H_0$ = 67.4 km s$^{-1}$ Mpc$^{-1}$, $\Omega_{vac}$ = 0.685 and $\Omega_m$ = 0.315 \citep{Ag20}.

\section{Search for Hybrid Morphology Radio Sources} 
\label{sec:style}
\subsection{MeerKAT absorption line survey (MALS)}
To search for hybrid morphology radio sources (HyMoRS), we used the MeerKAT absorption line survey \citep[MALS;][]{De24}. The MAL survey is observed at L-band (900--1670 MHz) using 391 telescope pointings at $\delta \leq+20^{\circ}$. The survey contains a catalog of 495,325 (240,321) radio sources detected at a signal-to-noise ratio (SNR) $>$5 over an area of 2289 deg$^2$ (1132 deg$^2$) at 1006 MHz (1381 MHz). The median spatial resolution is 12$\arcsec$ (8$\arcsec$). The median RMS noise away from the pointing center is 25 $\mu$Jy beam$^{-1}$ (22 $\mu$Jy beam$^{-1}$). In comparison with NVSS and FIRST at 1.4 GHz, the accuracy in the flux density scale and astrometry for sources in MALS is better than 6\% (15\% scatter) and 0.8$\arcsec$, respectively. 

\subsection{Identification of HyMoRS}
\label{subsec:sample}
We searched for HyMoRSs using the radio maps from MALS-DR1 \& DR2. The sources presented in this paper were selected based on the following criteria:

\begin{enumerate}[label=(\roman*)]
\item A detailed manual visual inspection of extended radio sources from MALS-DR1 \& DR2 was performed to identify possible candidates that show hybrid morphology. Sources were classified as HyMoRSs if they showed a clear FR I morphology on one side of the host galaxy and FR II morphology on the other. The morphology was further verified using radio maps from the Rapid ASKAP Continuum Survey mid-band data (RACS-mid; \citealt{askapm1}). Sources with ambiguous morphology or overlaps with unrelated nearby structures were excluded after visual assessment. The HyMoRS reported in the current paper are identified in the southern sky, where no extensive searches for HyMoRS have been carried out previously. Consequently, our cross-matching with previously reported HyMoRS in the literature yielded no overlaps. Therefore, no sources were excluded from the HyMoRS sample in the present study. 
Finally, thirty-six potential HyMoRS candidates were identified. At this stage, we also measured the integrated and core flux densities, along with the largest angular size (LAS), at the $3\sigma$ local noise level in the radio map.
    
\item Host galaxies were identified by overlaying radio images from MALS with optical and infrared data from the 9th data release from the Dark Energy Spectroscopic Instrument (DESI) Legacy Imaging survey \citep[DESI LS DR9;][]{Sc21} and AllWISE \citep{cu12, Cu13} survey, respectively. Hosts were considered reliable when the radio core coincided with an optical or IR galaxy. Mid-infrared measurements were obtained by cross-matching the newly discovered HyMoRS sample with the AllWISE source catalog. Redshift information (spectroscopic or photometric) was retrieved by cross-matching with updated catalogs \citep{Ahu20, Barrows21, Du22, glade22, flesch23} within a $2\arcsec$ radius.
\end{enumerate}

This selection process resulted in a refined and robust sample of thirty-six HyMoRS suitable for further statistical and environmental analysis. This is the largest sample of HyMoRS discovered in the southern sky. Out of thirty-six sources, twenty-five sources are detected at both frequencies of the MAL survey, i.e., at 1006 and 1381 MHz, whereas eleven sources are detected only at 1006 MHz and not at 1381 MHz.

\begin{table*}[h]
\movetabledown=6.6cm
\begin{rotatetable*}
\begin{scriptsize}
 \begin{center}
\caption{Radio properties of HyMoRSs in the present work: Col. (1): Serial number. Col. (2): Source name. Col. (3) and Col. (4): RA and DEC of host galaxy position. Col. (5): Redshift of host galaxy. Col. (6): Redshift reference. Col. (7) and Col. (8): Integrated radio flux densities at 1.0 GHz and 1.4 GHz. Col. (9): Integrated spectral indices. Col. (10) and Col. (11): Core radio flux densities at 1.0 GHz and 1.4 GHz. Col. (12): Core spectral indices. Col. (13): Total angular extent in arcminute. Col. (14) and Col. (15): FR index of FR I and FR II sides. Col. (16): Projected linear size (D). Col. (17) and (18): Total radio power at 1.0 GHz and 1.4 GHz. Col. (19): Core radio power at 1.4 GHz. Col. (20): Core dominance factor at 5 GHz. Col. (21): Doppler factor. Col. (22): Inclination angle.}
\label{tab:table1}
\begin{tabular}{cccccccccccccccccccccc}
    \hline
    \hline
No. &Source &RA &DEC &$z$ &$z_{ref}$ &$S_{1}$ &$S_{1.4}$ &$\alpha_{int}$ &$S_{1c}$ &$S_{1.4c}$ &$\alpha_{core}$ &LAS &$f_{FRI}$ &$f_{FRII}$ &$D$ &$P_{1}$ &$P_{1.4}$ &$P_{c}$ &CDF &$\delta$ &$\phi$ \\
~ &~ &(h:m:s) & (d:m:s) &~ &~ &(mJy) &(mJy) &~ &(mJy) &(mJy) &~ &$'$ &~ &~ &(kpc) &(\( \mathrm{W\,Hz^{-1}} \)) &(\( \mathrm{W\,Hz^{-1}} \)) &(\( \mathrm{W\,Hz^{-1}} \)) &~ &~ &($^\circ$) \\
(1) &(2) &(3) &(4) &(5) &(6) &(7) &(8) &(9) &(10) &(11) &(12) &(13) &(14) &(15) &(16) &(17) &(18) &(19) &(20) &(21) &(22) \\
    \hline
1&J0016--1208 &00:16:01.85 &--12:08:10.6 &0.988 & (i) &~13 &10 &0.71 &3.20 &2.98 &0.21 &0.87 &1.20 &1.80 &428 &25.34 &25.23 &24.85 &0.77 &1.57 &27 \\
2&J0039--0152 &00:39:22.57 &--01:52:14.4 &0.153 &(ii) &118 &89 &0.83 &4.20 &3.64 &0.43 &1.88 &1.30 &1.50 &308 &24.79 &24.67 &23.30 &0.06 &0.42 &58 \\
3&J0125+1441 &01:25:11.97 &+14:41:41.3 &0.499 &(ii) &~15 &12 &0.61 &4.05 &3.85 &0.15 &1.43 &1.40 &2.40 &541 &24.90 &24.81 &24.39 &0.96 &1.39 &29 \\
4&J0153--4837 &01:53:52.77 &--48:37:03.1 &0.323 & (i) &~11 &9 &0.69 &0.98 &1.30 &--0.82 &1.34 &1.30 &1.90 &387 &24.41 &24.31 &23.65 &1.18 &2.16 &22 \\
5&J0217--3846 &02:17:59.33 &--38:46:40.5 &0.724 & (i) &~20 &14 &1.01 &5.76 &4.90 &0.48 &0.97 &1.30 &1.70 &433 &25.22 &25.07 &24.74 &0.63 &1.19 &32 \\
6&J0335--5212 &03:35:45.02 &--52:12:07.9 &0.555 & (i) &~25 &19 &0.91 &1.67 &1.71 &--0.07 &1.36 &0.89 &1.50 &540 &25.15 &25.02 &24.17 &0.27 &0.98 &36 \\
7&J0339--2228 &03:39:56.38 &--22:28:56.2 &0.276 & (i) &~14 &11 &0.68 &0.68 &0.61 &0.31 &0.96 &1.40 &1.80 &248 &24.36 &24.26 &23.05 &0.10 &0.45 &55 \\
8&J0400--3616 &04:00:22.39 &--36:16:11.7 &0.604 & (i) &~17 &11 &1.28 &4.09 &3.80 &0.22 &1.42 &1.50 &1.60 &587 &24.96 &24.77 &24.54 &0.77 &1.26 &31 \\
9&J0407+1628 &04:07:46.25 &+16:28:45.0 &-- & -- &~61 &40 &1.25 &9.64 &8.84 &0.26 &1.69 &1.40 &2.40 &-- &-- &-- &-- &0.42 &-- &-- \\
10 &J0411--6402 &04:11:04.18 &--64:02:43.0 &0.143 &(iii) &~33 &27 &0.60 &-- &-- &0.25 &0.74 &1.30 &1.80 &114 &24.20 &24.11 &-- &-- &-- &-- \\
11 &J0509--3621 &05:09:25.53 &--36:21:05.5 &0.427 & (i) &~~7 &5 &0.98 &0.79 &0.67 &0.49 &1.49 &1.30 &1.50 &514 &24.39 &24.25 &23.44 &0.19 &0.57 &48 \\
12 &J0804+0812 &08:04:48.14 &+08:12:52.2 &0.233 &(ii) &~26 &19 &0.88 &8.74 &8.04 &0.25 &1.94 &1.50 &2.40 &446 &24.48 &24.35 &24.03 &1.15 &1.10 &33 \\
13 &J0833+0458 &08:33:52.25 &+04:58:22.4 &0.265 & (i) &162 &123 &0.82 &-- &-- &0.25 &1.75 &1.50 &1.60 &442 &25.39 &25.27 &-- &-- &-- &-- \\
14 &J0835+0436 &08:35:01.55 &+04:36:26.5 &0.761 & (i) &~~10 &8 &0.76 &0.15 &0.14 &0.39 &0.72 &1.50 &2.10 &329 &25.01 &24.89 &23.24 &0.03 &0.36 &63 \\
15 &J0944--3154 &09:44:46.42 &--31:54:39.8 &-- & -- &~62 &51 &0.59 &20.20 &18.50 &0.26 &1.36 &1.30 &2.10 &-- &-- &-- &-- &1.05 &-- &-- \\
16 &J0948--3110 &09:48:01.84 &--31:10:56.3 &0.190 &(iii) &~16 &12 &0.76 &2.35 &1.99 &0.49 &1.14 &1.10 &2.10 &223 &24.11 &24.00 &23.22 &0.26 &0.58 &48 \\
17 &J0953--3253 &09:53:38.05 &--32:53:52.0 &-- & -- &~11 &9 &0.64 &1.33 &1.28 &0.11 &1.49 &1.20 &2.60 &-- &-- &-- &-- &0.39 &-- &-- \\
18 &J0954--0003 &09:54:36.07 &--00:03:34.3 &0.390 &(ii) &~15 &12 &0.60 &3.43 &3.21 &0.20 &1.16 &1.20 &1.80 &379 &24.70 &24.61 &24.09 &0.73 &1.13 &33 \\
19 &J1128--0638 &11:28:53.54 &--06:38:52.4 &0.183 &(ii) &~62 &43 &1.06 &6.80 &6.47 &0.15 &1.91 &0.90 &1.50 &364 &24.64 &24.49 &23.73 &0.33 &0.73 &42 \\
20 &J1243--0100 &12:43:38.21 &--01:00:30.0 &1.336 & (i) &~13 &8 &1.37 &0.75 &0.69 &0.24 &1.29 &1.10 &2.30 &668 &25.29 &25.09 &24.44 &0.14 &0.89 &38 \\
21 &J1313--2755 &13:13:26.69 &--27:55:28.9 &0.197 &(iii) &~41 &31 &0.85 &3.84 &3.68 &0.13 &1.91 &0.80 &1.90 &385 &24.54 &24.41 &23.55 &0.28 &0.68 &44 \\
22 &J1338--2926 &13:38:41.49 &--29:26:17.0 &-- & -- &~19 &12 &1.39 &1.29 &1.27 &0.05 &1.11 &1.50 &2.20 &-- &-- &-- &-- &0.23 &-- &-- \\
23 &J1348--2409 &13:48:33.63 &--24:09:08.2 &0.500 &(iv) &179 &127 &1.03 &34.80 &30.00 &0.44 &1.32 &1.30 &1.90 &498 &25.91 &25.76 &25.23 &0.40 &1.29 &30 \\
24 &J1445--1135 &14:45:11.89 &--11:35:43.5 &-- & -- &~~7 &4 &1.38 &2.11 &1.93 &0.27 &0.99 &1.50 &1.80 &-- &-- &-- &-- &0.94 &-- &-- \\
25 &J1606--1844 &16:06:40.47 &--18:44:44.3 &0.286 &(ii) &208 &162 &0.74 &21.30 &18.40 &0.43 &1.32 &1.49 &1.60 &351 &25.57 &25.46 &24.55 &0.19 &0.88 &38 \\
26 &J1626--3633 &16:26:38.17 &--36:33:19.8 &0.043 &(iii) &367 &279 &0.81 &-- &-- &0.25 &1.98 &1.43 &2.17 &105 &24.20 &24.08 &-- &-- &-- &-- \\
27 &J1959--1852 &19:59:19.00 &--18:52:20.7 &0.610 &(v) &~48 &39 &0.62 &18.30 &17.20 &0.18 &1.57 &1.47 &1.80 &654 &25.56 &25.47 &25.21 &1.54 &2.09 &23 \\
28 &J2130--2801 &21:30:31.92 &--28:01:21.1 &-- & -- &~~8 &6 &0.64 &-- &-- &0.25 &0.94 &1.59 &1.32 &-- &-- &-- &-- &-- &-- &-- \\
29 &J2137--5405 &21:37:18.69 &--54:05:41.3 &0.404 & (i) &~~9 &7 &0.93 &3.68 &3.24 &0.38 &1.38 &1.52 &2.02 &462 &24.45 &24.32 &24.10 &1.28 &1.09 &34 \\
30 &J2223--4543 &22:23:33.12 &--45:43:03.8 &0.369 &(iii) &~12 &9 &1.00 &2.77 &2.52 &0.28 &1.58 &1.16 &1.60 &501 &24.51 &24.36 &23.92 &0.62 &0.92 &37 \\
31 &J2226+1310 &22:26:10.85 &+13:10:26.2 &0.214 & (i) &~11 &-- &-- &2.61 &2.29 &0.39 &1.15 &1.40 &1.70 &247 &24.04 &-- &23.40 &0.56 &0.73 &42 \\
32 &J2231--1406 &22:31:02.20 &--14:06:13.4 &0.407 & (i) &~12 &10 &0.59 &3.24 &3.15 &0.08 &1.07 &1.50 &1.90 &360 &24.63 &24.54 &24.13 &1.13 &1.34 &30 \\
33 &J2301--5334 &23:01:44.26 &--53:34:48.5 &0.240 & (i) &~66 &49 &0.90 &14.00 &13.50 &0.11 &1.82 &1.10 &2.30 &426 &24.91 &24.78 &24.29 &0.76 &1.20 &32 \\
34 &J2319--4632 &23:19:42.10 &--46:32:59.6 &0.877 & (i) &~27 &21 &0.73 &0.65 &0.61 &0.19 &1.70 &1.30 &2.10 &811 &25.57 &25.46 &24.06 &0.06 &0.60 &47 \\
35 &J2336--3735 &23:36:04.30 &--37:35:37.7 &-- & -- &~~6 &5 &0.70 &0.77 &0.66 &0.46 &1.30 &1.10 &1.80 &-- &-- &-- &-- &0.25 &-- &-- \\
36 &J2357--0725 &23:57:06.19 &--07:25:39.2 &0.536 &(ii) &~~10 &7 &1.13 &0.50 &0.48 &0.12 &0.96 &1.30 &2.50 &376 &24.68 &24.51 &23.55 &0.15 &0.58 &48 \\
    \hline
\end{tabular}
\end{center}
Redshift references-- (i): \citet{Du22}; (ii): \citet{Ahu20}; (iii): \citet{glade22}; (iv): \citet{flesch23}; (v): \citet{Barrows21};
\end{scriptsize}
\end{rotatetable*}
\end{table*}

\section{Results} 
\label{sec:result}
\begin{figure*}
\vbox{
\centerline{
\includegraphics[width=20.0cm, origin=c]{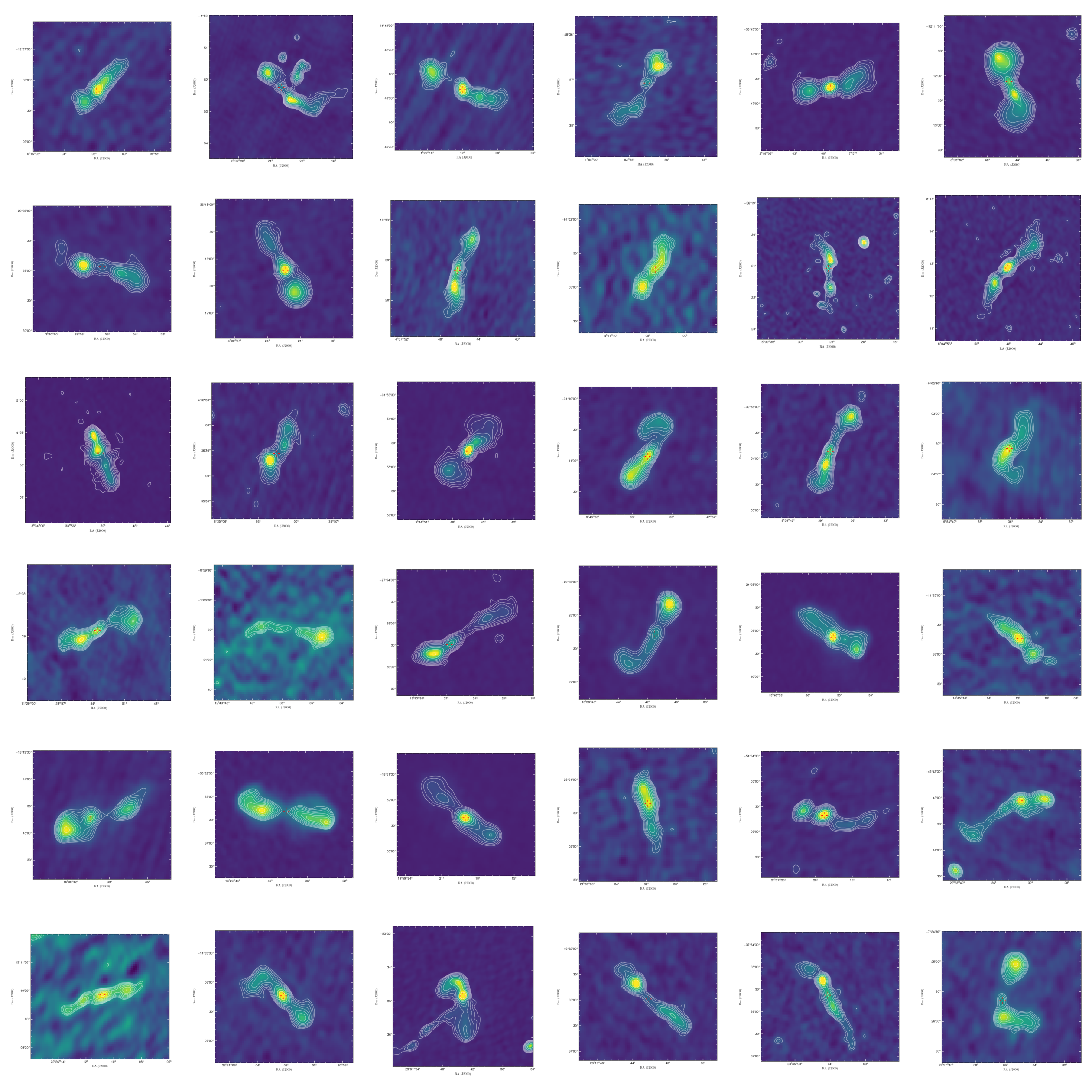}
}
}
\caption{Contour-overlaid radio images of all thirty-six HyMoRSs identified in this study. The red cursor marks indicate the positions of the host galaxies, and the white contours correspond to MALS 1000 MHz images. The contours are plotted at ten logarithmically spaced levels, starting from 3$\sigma$, where $\sigma$ denotes the local background RMS of each image. 
}
\label{fig:fullsample}
\end{figure*}

The list of thirty-six identified HyMoRSs using the MAL survey is tabulated in Table \ref{tab:table1}. All the presented HyMoRSs have IR counterparts near the center of the sources. Figure \ref{fig:sample} displays a representative sample of four HyMoRSs, with the optical host galaxy of each source marked by a black plus sign ($+$) near the center. The components labeled F1 and F2 correspond to the FR I and FR II lobes, respectively, illustrating the hybrid morphology (see Figure \ref{fig:sample} for details). Contour-overlaid images of all thirty-six HyMoRSs in our sample are shown in Figure \ref{fig:fullsample}. In Section~\ref{subsec:rad_props}, we present the radio properties of the newly identified HyMoRS sample, and in Section~\ref{subsec:host_prop}, we discuss the host galaxy properties of these sources.

\subsection{Radio Properties}
\label{subsec:rad_props}
\subsubsection{Estimation of FR-Index ($f_{\rm FR}$)}
Since the selection of HyMoRSs in this work was based on the manual visual inspection method, we computed the $f_{\rm FR}$ index \citep{kr12} to provide a quantitative evaluation of their radio morphology. This index is derived from the original classification by \citet{Fa74} and is defined as:  

\begin{equation}	
     f_{\rm FR}=\frac{2p}{l}+0.5
    \label{equ:FR}
\end{equation}  
where $p$ represents the angular distance of the brightest pixel from the core, and $l$ is the angular extent of the individual FR lobe. According to the classification scheme of \citet{kr12} and \citet{Ka17}, HyMoRS sources exhibit an FR index of $f_{\rm FR} > 1.5$ (FR II) on one side and $0.5 < f_{\rm FR} < 1.5$ (FR I) on the other side.  

There are some limitations of using this definition. For instance, if the observations fail to capture the extended emission from the faint outer regions of the lobe, the index value might be overestimated. Additionally, this method works best when a lobe has a distinct brightness peak; however, if the brightness remains fairly uniform as the distance from the core increases, the index may not accurately reflect the lobe's characteristics. Therefore, rather than serving as a strict classification, the $f_{\rm FR}$ index should be considered an indicative measure \citep{Ka17}. We calculate the $f_{\rm FR}$ index for each lobe in each HyMoRS candidate. 

\begin{figure}
\vbox{
\centerline{
\includegraphics[width=9.5cm, origin=c]{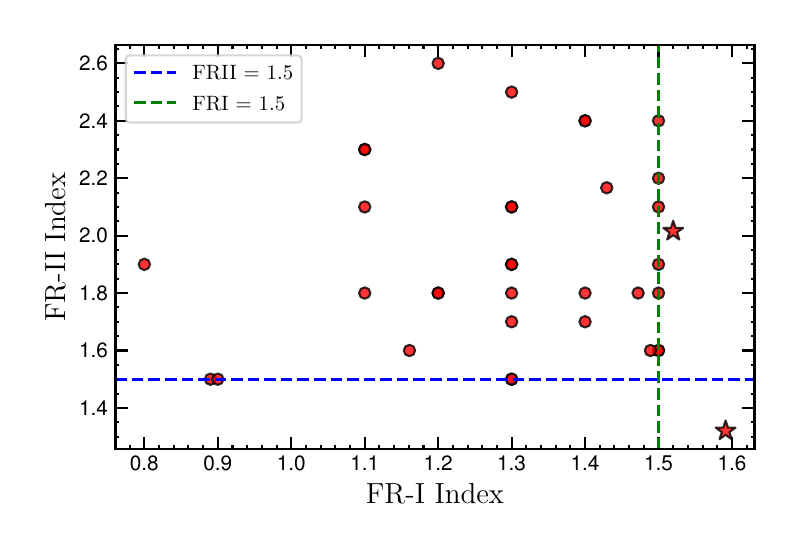}
}
}
\caption{Distribution of FR-Index along FR I and FR II sides. Here, the red star points denote the HyMoRSs which dominate the FR-index definition whereas sources, represented with red points follow the FR-index definition.}
\label{fig:FR_index}
\end{figure}

The FR index ($f_{\rm FR}$) for the FR I side ($f_{\rm FRI}$) and FR II side ($f_{\rm FRII}$) are presented in columns 14 and 15 of Table \ref{tab:table1}. Figure \ref{fig:FR_index} shows an FR diagnostic diagram that illustrates the distribution of the FR index for the FR I and FR II sides. In the plot, almost all of the HyMoRS candidates (represented by red points) follow the FR index diagnosis criterion, showing $f_{\rm FR} > 1.5$ (FR II) on one side and $0.5 < f_{\rm FR} < 1.5$ (FR I) on the other side. However, two sources, J2130--2801 and J2137--5405 (marked as red stars in Figure \ref{fig:FR_index}), do not satisfy this $f_{\rm FR}$ criterion. Similar exceptions have also been reported in previous studies on HyMoRS \citep{Pa22, Ka17}, where some sources did not adhere to the $f_{\rm FR}$ classification criterion.

The FR-Index ($f_{\rm FR}$) for FR I lobes of the HyMoRSs presented here ranges from 0.8 to 1.5, with an average and median value of 1.3. Meanwhile, the FR-Index ($f_{\rm FR}$) for FR II lobes of HyMoRSs varies between 1.5 and 2.6, with a mean value of 1.93 and a median of 1.85. The average values of the FR-Index for FR I and FR II lobes are comparable with that of 3CRR sources which have an average of 1.13 for FR I lobes and 2.13 for FR II lobes \citep{La83,Ka17}. The average FR-Indexes for FR I lobes and FR II lobes for HyMoRS presented in \citet{Ka17} were 1.18 and 2.14 and in \citet{Pa22} the average FR-Indexes for FR I lobes and FR II lobes were 1.28 and 2.04, respectively, which are also close to the average FR-Index of our sample. We excluded two sources (J2130--2801 and J2137--5405) to avoid bias in the comparison of statistical properties with earlier catalogs, where they also restricted those sources which don't follow the $f_{\rm FR}$ classification criterion while calculating the mean and median of their samples.

\subsubsection{Spectral Index ($\alpha$)}
\begin{figure}
\vbox{
\centerline{
\includegraphics[width=9.5cm, origin=c]{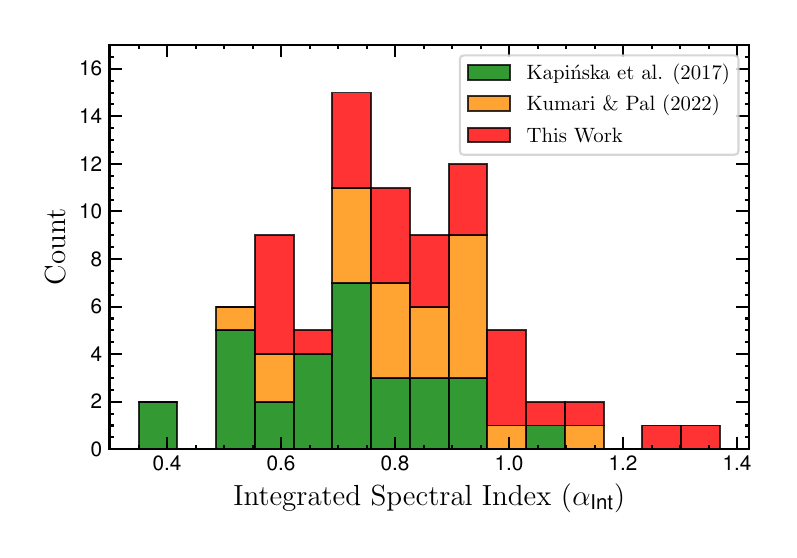}
}
}
\caption{Histogram showing the spectral-index distribution of HyMoRS reported in the current paper, along with \citet{Pa22} and \citet{Ka17}.}
\label{fig:spectral_index}
\end{figure}

We computed the two-point core (\(\alpha_{\text{core}}\)) and integrated (\(\alpha_{\text{Int}}\)) spectral indices of HyMoRSs between 1.0 GHz and 1.4 GHz using the measured core and integrated flux densities, assuming a power-law spectrum (\( S \propto \nu^{-\alpha} \), where $\alpha$ is the spectral index). The estimated integrated and core spectral indices are tabulated in columns 9 and 12 of Table \ref{tab:table1}. Figure \ref{fig:spectral_index} presents the distribution of the integrated spectral index of HyMoRSs presented in the current paper along with the HyMoRSs presented in \citet{Pa22} and \citet{Ka17}. The integrated spectral index (\(\alpha_{\text{Int}}\)) for sources presented in the current paper lies between 0.59 to 1.39 with a mean spectral index of 0.87 (median = 0.82). The integrated spectral index of all the 36 HyMoRSs shows steep radio spectra (\(\alpha_{\text{Int}}>0.5\)), which is the common property of lobe-dominated radio galaxies.  The mean spectral indices in \citet{Pa22} and \citet{Ka17} are 0.82 and 0.70, respectively. So, the sources presented in the current paper are steeper than previously discovered HyMoRS in \citet{Pa22} and \citet{Ka17}. The core spectral index ($\alpha_{\text{core}}$) for HyMoRSs in this study ranges from --0.82 to 0.49, with a mean of 0.23 and a median of 0.25. This indicates a predominantly flat radio spectrum ($\alpha_{\text{core}} < 0.5$), as expected for radio cores.

\subsubsection{Spectral Index Asymmetry between the lobes}
\begin{figure}
\vbox{
\centerline{
\includegraphics[width=9.5cm, origin=c]{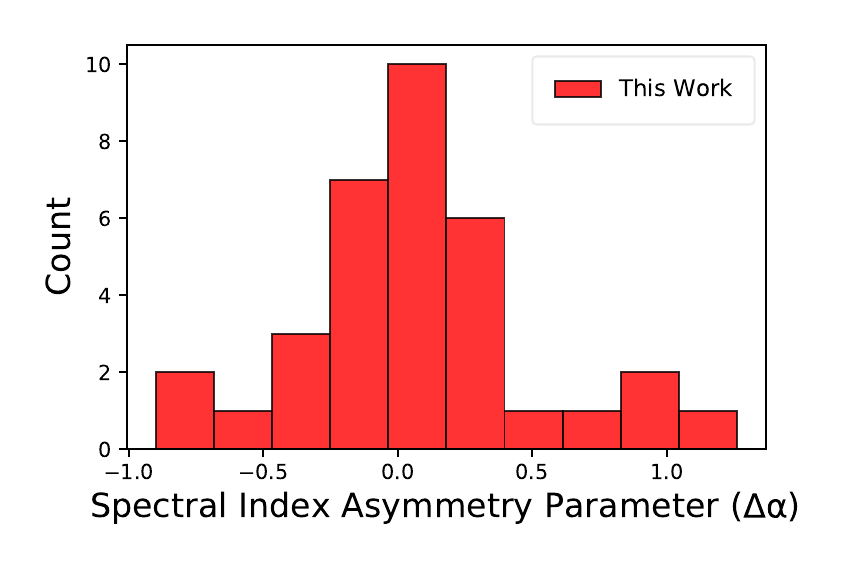}
}
}
\caption{Distribution of the spectral index asymmetry parameter (\(\Delta\alpha\)) between the two lobes of HyMoRSs in the current paper.}
\label{fig:delta_alpha_hist}
\end{figure}

To examine whether the two sides of our sources exhibit systematic spectral differences, we introduced the \textit{spectral index asymmetry parameter} ($\Delta\alpha$). This parameter is defined as
\begin{equation}
    \Delta \alpha = \alpha_{\mathrm{FRI}} - \alpha_{\mathrm{FRII}},
\end{equation}
where $\alpha_{\mathrm{FRII}}$ is the spectral index measured for the lobe with FR II-type morphology and $\alpha_{\mathrm{FRI}}$ is that of the opposite lobe with FR I-type morphology. Figure~\ref{fig:delta_alpha_hist} presents the distribution of the spectral index asymmetry parameter (\(\Delta\alpha\)) between the two lobes of HyMoRSs in the current work. For our HyMoRS sample, the distribution of $\Delta \alpha$ has a mean of 0.09, a median of 0.08, and a standard deviation of 0.45. Since our goal was to test whether the spectral indices of the two lobes are similar, we restricted our analysis to paired tests that directly assess differences within the same source, as the two lobes belong to the same galaxy and are therefore not independent measurements. Accordingly, we applied three complementary tests: a paired \textit{t}-test, a Wilcoxon signed-rank test, and a paired permutation test. All three tests yielded $p$-values well above the conventional significance threshold (paired \textit{t}-test: $p = 0.250$; Wilcoxon: $p = 0.179$; permutation: $p = 0.251$), consistently indicating that the observed mean difference is not statistically significant ($p>0.05$). The effect size is small (Cohen’s $d = 0.20$), further suggesting that any systematic offset, if present, is negligible compared to the source-to-source scatter. Taken together, these results demonstrate that the two lobes of HyMoRS are statistically consistent with having similar spectral indices. This implies that despite their hybrid morphology, the lobes undergo broadly comparable particle acceleration and radiative loss processes, with the observed scatter likely arising from random or local environmental effects rather than an intrinsic asymmetry.

\subsubsection{Radio Power}
We estimated the core and total radio power ($P_{\nu,Core}$ and $P_{\nu,Int}$) for the presented HyMoRS using Eq. \ref{equ:LogL}:

\begin{equation}	
    P_{\nu}=4\pi{D_{L}}^{2}S_{\nu}(1+z)^{\alpha-1}
    \label{equ:LogL}
\end{equation}
where $z$ is the redshift, $D_L$ is the luminosity distance to the source, $S_{\nu}$ represents the flux density at the frequency $\nu$, and $\alpha$ denotes the two-point spectral index calculated for the core and integrated emission of HyMoRSs between 1.0 GHz and 1.4 GHz. 

\begin{figure}
\vbox{
\centerline{
\includegraphics[width=9.5cm, origin=c]{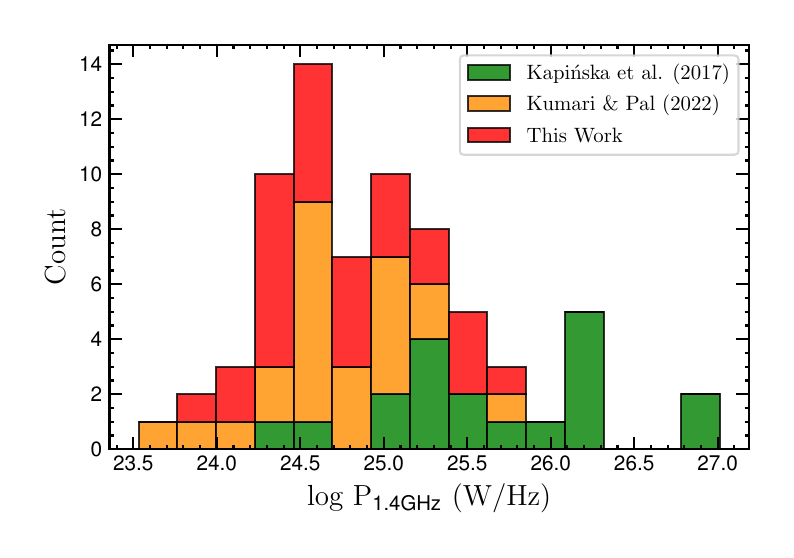}
}
}
\caption{Histogram showing distribution of total radio power at 1.4 GHz for HyMoRSs in the current paper along with \citet{Pa22} and \citet{Ka17}.}
\label{fig:p1.4}
\end{figure}

The integrated and core flux densities at 1.0 GHz and 1.4 GHz for the HyMoRS in the current article are presented in columns~7, 8, 10 and 11 of Table \ref{tab:table1}. The corresponding total and core radio powers ($\rm P_{1.4,Int}$ and $\rm P_{1.4,Core}$ in W Hz$^{-1}$) of the sources at 1.4 GHz are presented in columns 18 and 19 of Table \ref{tab:table1}. Figure \ref{fig:p1.4} shows the distribution of $\rm P_{1.4,Int}$ for the sources presented in this study, along with those from \citet{Pa22} and \citet{Ka17}.%, while Figure \ref{fig:P-z} illustrates its variation with redshift
 The grey dashed line in Figure \ref{fig:P-z} represents the minimum luminosity at different redshifts, corresponding to the faintest object in the present HyMoRS sample, which has a flux density of 4.40 mJy, assuming an integrated spectral index of 0.75. The core radio power at 1.4 GHz ($\rm \log~P_{1.4,Core}$[W Hz$^{-1}$]) for the present HyMoRS sample ranges from 23.05 to 25.23, with a mean and median value of 24.03 and 24.08, respectively. The total radio power ($\rm \log~P_{1.4,Int}$[W Hz$^{-1}$]) for HyMoRSs in the present paper varies from 23.99 to 25.75, with a mean value of 24.73 and a median of 24.64. This suggests that the HyMoRS discussed in this paper have radio luminosities near the FR I/II division line (\(P_{150} \sim 10^{26} \, \text{W Hz$^{-1}$}\)).

\subsubsection{Projected Linear Size ($D$)}
We determined the total angular extent, or largest angular sizes (LAS), of the discovered HyMoRSs using radio maps from the MALS survey at a $3\sigma$ significance level. Using the measured LAS, we computed the projected linear size as  
\begin{equation}  
D = \frac{\theta \times D_{co}}{(1+z)}  
\label{equ:D}
\end{equation}  
where $\theta$ is the LAS of the source, $z$ is the redshift, and $D_{co}$ is the comoving distance of the source. We presented the linear size estimates of the HyMoRSs in the present paper in column 16 of Table \ref{tab:table1}. HyMoRSs in the present paper have projected linear sizes ranging from 105 kpc to 811 kpc with a mean and median value of 418 kpc and 426 kpc, respectively. All the sources presented in this study have linear sizes of less than 700 kpc, except for one source, J2319--4632, which is classified as a giant radio galaxy \citep[e.g.][]{Bh24,Ku24,Manik25} with a linear size of 811 kpc. 

\subsubsection{Orientation of HyMoRS}
\label{subsubsec:orientation}
We analyzed the orientation of HyMoRSs to understand the role of Doppler boosting and inclination angle in their observed morphology. To do so, we estimated the Doppler factor following the approach of \citet{gio01}, which is based on the correlation between core and total radio power in radio galaxies, which has been widely applied to radio galaxies:

\begin{equation}
   \rm log~P_{ci5}=0.62~log~P_{t}+8.41
\label{equ:Pc}    
\end{equation} 
where $\rm P_{ci5}$ is the intrinsic core luminosity at 5 GHz, estimated assuming $\rm \gamma = 5$, and $\rm P_t$ represents the total radio luminosity at 408 MHz. Here, both $\rm P_{ci5}$ and $\rm P_t$ are in W Hz$^{-1}$. Although this relation was calibrated primarily on a sample dominated by FR I sources, \citet{lara04} later confirmed its validity for a larger sample of large angular size FR I and FR II radio galaxies. The same calibration was also previously employed by \citet{Ce13} for the estimation of Doppler factors for HyMoRSs.

The observed core radio luminosity at 5 GHz, $\rm P_{c,o}$, is affected by Doppler boosting and is related to $\rm P_{ci5}$ by:

\begin{equation}
\rm P_{c,o} = P_{ci5} \delta^{2+\alpha_{core}}
\label{equ:delta}    
\end{equation} 
where $\rm P_{c,o}$ is the observed core radio luminosity at 5 GHz, $\delta$ is the Doppler factor, and $\alpha_{core}$ is the core spectral index.

Rewriting Equation \ref{equ:delta} to estimate $\delta$: 

\begin{equation}
\rm \delta = \left( \frac{P_{c,o}}{P_{ci5}} \right)^{\frac{1}{2+\alpha_{core}}} 
\label{equ:delta1}    
\end{equation}
%\delta = \sqrt{\frac{P_{c,o}}{P_{ci5}}}
Since all sources are detected from the 1006 MHz MALS survey, to estimate the integrated radio flux density at 408 MHz and core flux density at 5 GHz we use the following formula:

\begin{equation}
\rm S_\nu(\nu_\mathrm{rest}) = S_\nu(\nu_\mathrm{obs}) \left(\frac{\nu_\mathrm{rest}}{\nu_\mathrm{obs}}\right)^{-\alpha}
\label{equ:nurest}    
\end{equation}
where $\rm S_{\nu_{rest}}$ denotes the rest-frame flux density at $\rm \nu_{rest}$, while $\rm S_{\nu_{obs}}$ represents the observed flux density at $\rm \nu_{obs}$ for both the core and extended radio emissions of the source. For the estimation of the integrated radio flux density at 408 MHz, we used the estimated integrated spectral index and for the estimation of the core flux density at 5 GHz, we used the estimated core spectral index. 

Moreover, the Doppler factor ($\delta$) is given by:

\begin{equation}
\delta = \frac{1}{\gamma (1 - \beta \cos\phi)}
\label{equ:deltagamma}    
\end{equation}
where $\gamma$ is the Lorentz factor, $\beta = \frac{v}{c}$ is the velocity in units of $c$, and $\phi$ is the inclination angle. Rearranging equation \ref{equ:deltagamma} to express $\cos\phi$ in terms of the Lorentz factor $\gamma$ and the Doppler factor $\delta$, and substituting $\beta = \sqrt{1 - \frac{1}{\gamma^2}}$, we obtain the inclination angle $\phi$ by taking the inverse cosine:

\begin{equation}
\phi = \cos^{-1} \left[ \frac{1}{\sqrt{1 - \frac{1}{\gamma^2}}} \left(1 - \frac{1}{\gamma \delta} \right) \right]
\label{equ:inclination}    
\end{equation}

\begin{figure}
\vbox{
\centerline{
\includegraphics[width=9.5cm, origin=c]{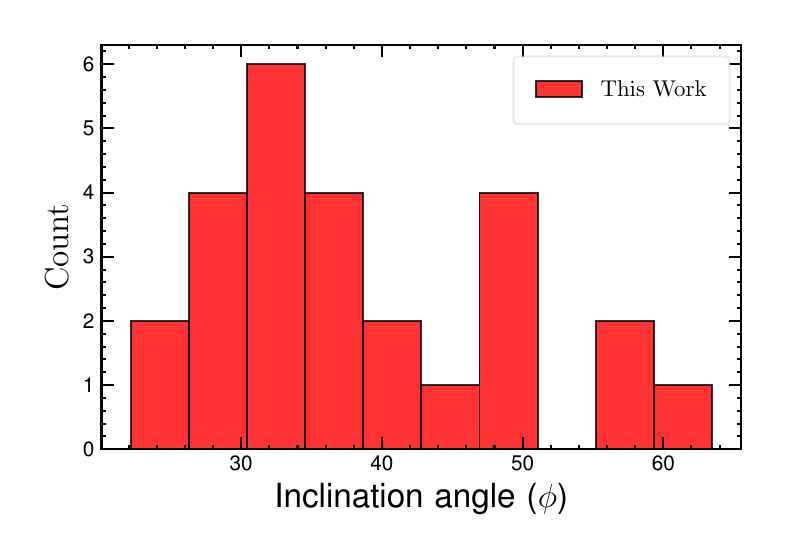}
}
}
\caption{Distribution of inclination angle (\(\phi\)) of HyMoRSs in the present work.}
\label{fig:inc}
\end{figure}

For the present HyMoRS sample, we estimated the inclination angle assuming $\rm \gamma = 5$ and using estimated Doppler factor ($\delta$) from Equation \ref{equ:delta1}. We present the estimates of $\delta$ and $\phi$ in Columns 21 and 22 of Table \ref{tab:table1}. In Figure \ref{fig:inc}, we present the distribution of the estimated inclination angle (\(\phi\)) of HyMoRSs in the present work. The Doppler factor ($\delta$) for HyMoRSs ranges from 0.36 to 2.16 with a mean value of 1.0 and a median of 0.95. The estimated inclination angle for the present HyMoRS sample varies between 22$^\circ$ to 63$^\circ$ with a mean value of 38$^\circ$ and a median of 36$^\circ$. These values are consistent with expectations for unbeamed radio galaxies observed at moderate viewing angles. For HyMoRS specifically, VLBI studies by \citet{Ce13} found similarly low Doppler factors and large viewing angles, and showed no systematic connection between parsec-scale jet orientation and the FR I or FR II side. More recent work \citep{Ha20,Stroe22} has highlighted the dominant role of environmental asymmetries in shaping hybrid morphologies. Taken together, this suggests that while our estimates provide a useful comparative reference, they should be regarded as approximate indicators rather than precise determinations of the beaming properties of HyMoRS.

\subsubsection{Core Dominance Factor}
The \textit{core dominance factor} (\( CDF \)) is a key parameter in quantifying the relative strength of the core emission compared to the extended radio structure. A higher \( CDF \) value suggests a stronger, possibly Doppler-boosted core, while a lower \( CDF \) indicates that most of the emission originates from the extended structure. This is particularly useful in studying the hybrid morphology of radio sources, where one side may show stronger emission due to environmental or relativistic effects. It is defined as the ratio of the core flux density to the extended flux density:

\begin{equation}
CDF = \frac{S_{\text{core, 5GHz}}}{S_{\text{ext, 5GHz}}} = \frac{S_{\text{core, 5GHz}}}{S_{\text{int, 5GHz}} - S_{\text{core, 5GHz}}}
\end{equation}
where \( S_{\text{int, 5GHz}} \) is the integrated flux density at 5 GHz, \( S_{\text{core, 5GHz}} \) is the core flux density at 5 GHz, and \( S_{\text{ext, 5GHz}} \) represents the flux from the extended structure. Since the integrated flux at 5 GHz includes both the core and extended emission, we calculate the extended flux density by subtracting \( S_{\text{core, 5GHz}} \) from \( S_{\text{int, 5GHz}} \). 

To estimate the core and extended flux densities at 5 GHz from the 1 GHz flux density obtained from MALS images, we use Equation \ref{equ:nurest} along with a spectral index. Specifically, we use the estimated core spectral index for the core component and assume a spectral index of 0.75 for the extended emission. We present the estimated \( CDF \) in column 20 of Table \ref{tab:table1}. \( CDF \) for HyMoRSs in the present work varies between 0.03 to 1.54 with a mean value of 0.56 and a median of 0.41.

\subsection{Mid-infrared Properties of the Host Galaxies of HyMoRSs}
\label{subsec:host_prop}
The mid-infrared (MIR) spectrum of HyMoRS host galaxies provides valuable insights into the radiative efficiency of the central supermassive black hole. The surrounding dusty torus absorbs optical-ultraviolet radiation from the AGN accretion disk and re-emits it in the MIR bands. To study the host properties of HyMoRSs at mid-IR wavelengths, we used data from the ALLWISE all-sky IR survey. ALLWISE survey was performed using a space-based telescope WISE, observed in four bands (W1, W2, W3, and W4) corresponding to 3.4 $\mu$m, 4.6 $\mu$m, 12 $\mu$m, and 22 $\mu$m wavelengths with an angular resolution of 6.1$\arcsec$, 6.4$\arcsec$, 6.5$\arcsec$, and 12$\arcsec$, respectively. In the following, we discuss the MIR properties of the present HyMoRS sample as well as the previously discovered samples from \citet{Pa22} and \citet{Ka17}. In section \ref{subsubsec:IRcolor}, we discussed the classification of HyMoRSs based on their radiative efficiency using the WISE color-color diagram, in section \ref{subsubsec:IRluminosity}, we explore the MIR luminosities of HyMoRSs, in section \ref{subsubsec:Mstar}, we discussed the stellar mass estimates of HyMoRSs and in section \ref{subsubsec:SFR}, we studied the star formation rates of HyMoRSs.
\subsubsection{WISE Colors}
\label{subsubsec:IRcolor}
The WISE mid-infrared color--color plot is a useful tool to classify radio galaxies into different subclasses such as high-excitation radio galaxies (HERGs), low-excitation radio galaxies (LERGs), star-forming galaxies (SFGs), quasars, ultraluminous infrared radio galaxies (ULIRGs), and narrow-line radio galaxies (NLRGs). These subclasses occupy distinct regions in the WISE color--color plane:  
\begin{itemize}
    \item AGN/HERGs: $(W1 - W2) \geqslant 0.5$ and $(W2 - W3) < 5.1$
    \item LERGs (ellipticals): $(W1 - W2) < 0.5$ and $0 < (W2 - W3) < 1.6$
    \item Star-forming galaxies (SFGs): $(W1 - W2) < 0.5$ and $1.6 \leqslant (W2 - W3) < 3.4$
    \item ULIRGs: $(W1 - W2) < 0.5$ and $(W2 - W3) \geqslant 3.4$
\end{itemize}

\begin{figure*}
\vbox{
\centerline{
\includegraphics[width=12.0cm, origin=c]{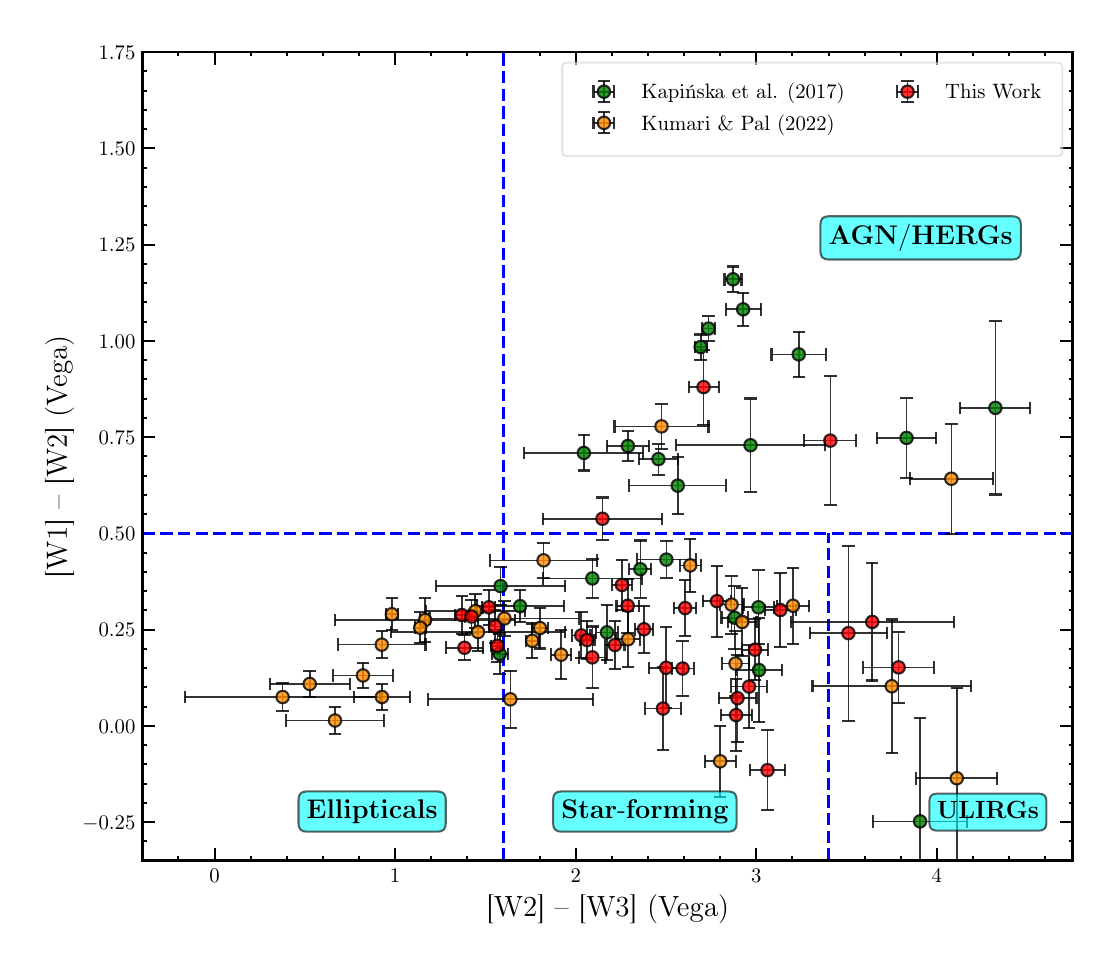}
}
}
\caption{WISE color--color plot of HyMoRS in the current paper, along with \citet{Pa22} and \citet{Ka17}.}
\label{fig:WISE_color}
\end{figure*}

Figure~\ref{fig:WISE_color} shows the WISE IR color--color distribution of the HyMoRSs presented in this work, along with sources reported by \citet{Pa22, Ka17}. Of the 36 HyMoRSs in our sample, 34 are detected in the ALLWISE survey. In the WISE color--color plot, 20 HyMoRS (59\%) fall in the SFG region, 7 HyMoRS (20\%) in the LERG (ellipticals) region, 4 HyMoRS (12\%) in the ULIRG region, and 3 HyMoRS (9\%) in the AGN/HERG region. Thus, the majority of HyMoRSs in our sample appear to be associated with Star-forming galaxies. However, a detailed spectroscopic analysis is required to confirm their true excitation class and disentangle possible contamination from dust-obscured star formation.

\subsubsection{Mid-infrared Luminosities}
\label{subsubsec:IRluminosity}
The MIR emission provides crucial insights into the physical properties of HyMoRS by tracing key astrophysical processes such as star formation, dust heating, and AGN activity. The emission in this wavelength range primarily originates from thermal reradiation by dust, which absorbs optical-ultraviolet photons from the accretion disk and re-emits them in the MIR bands. Using the measured WISE mid-infrared (MIR) band magnitudes and the redshifts of the HyMoRS host galaxies, we estimated the MIR luminosities at 3.4~$\mu$m, 4.6~$\mu$m, 12~$\mu$m, and 22~$\mu$m. These luminosities are presented in Columns 9, 10, 11, and 12 of Table~\ref{tab:ir_host}. The range of luminosities at 3.4 $\mu$m spans from $4.32 \times 10^{9}$ to $4.74 \times 10^{11}$ $L_\odot$, with a mean value of $6.12 \times 10^{10}$ $L_\odot$ and a median of $3.98 \times 10^{10}$ $L_\odot$. At 4.6 $\mu$m, the luminosity varies between $2.35 \times 10^{9}$ $L_\odot$ and $5.57 \times 10^{11}$ $L_\odot$, with a mean of $4.13 \times 10^{10}$ $L_\odot$ and a median of $2.12 \times 10^{10}$ $L_\odot$. Similarly, for 12 $\mu$m, the luminosities range from $3.24 \times 10^{8}$ to $6 \times 10^{11}$ $L_\odot$, with a mean of $4.28 \times 10^{10}$ $L_\odot$ and a median of $1.34 \times 10^{10}$ $L_\odot$. Finally, at 22 $\mu$m, the estimated luminosities lie between $4.36 \times 10^{8}$ and $1.06 \times 10^{12}$ $L_\odot$, with a mean of $1.04 \times 10^{11}$ $L_\odot$ and a median of $4.23 \times 10^{10}$ $L_\odot$.

\subsubsection{Stellar Masses}
\label{subsubsec:Mstar}
\begin{figure}
\vbox{
\centerline{
\includegraphics[width=9.5cm, origin=c]{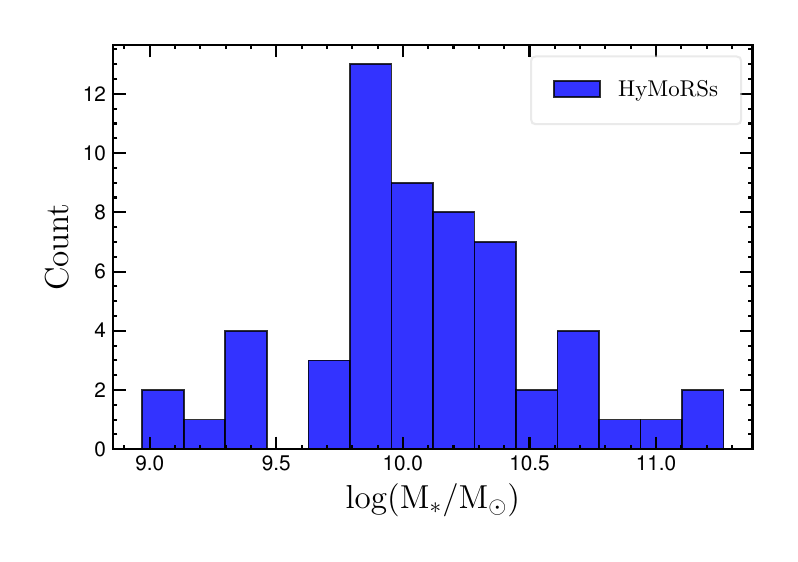}
}
}
\caption{Distribution of stellar masses (M$_{*}$) of combined sample of HyMoRS consists of sources with W1--W2 $\leq$ 0.5 from present work along with \citet{Pa22} and \citet{Ka17}. }
\label{fig:mstar}
\end{figure}

Stellar mass ($M_*$) is a crucial parameter for understanding galaxy evolution and jet activity cycles, including the restarted activity seen in HyMoRS. The MIR ALLWISE W1 and W2 photometry offers an effective way to estimate stellar mass content within the galaxy by tracing the Rayleigh-Jeans limit of blackbody emission from stars with temperatures of 2000 $K$ and above \citep{Jarrett13}.
\begin{equation}
\log_{10}(M_*/L_{W1}) = -1.96 \times (W1 - W2) - 0.03
\end{equation}
where $M_*$ is the stellar mass in unit of solar mass ($M_\odot$) and $L_{W1}$ is the luminosity of the MIR W1 band (3.4 $\mu$m) in solar luminosity ($L_\odot$). The W1--W2 color value remains relatively constant and is not influenced by the age of the stellar population or the mass function \citep{pahre2004, jarrett2011, cluver2014}. However, it can be affected by redshift, as demonstrated in studies by \citet{donley2012, assef2013}. We estimated the stellar masses and star formation rates (SFRs) for the HyMoRS sample presented in this study, along with the sources from \citet{Pa22} and \citet{Ka17}. The stellar mass estimates are provided in Column~12 of Table~\ref{tab:ir_host}. The stellar mass (\(M_*\)) of the combined HyMoRS sample varies between $9.35 \times 10^{8} M_{\odot}$ and 1.85$\times 10^{11}~M_{\odot}$, with a mean value of 2.29$\times 10^{10}~M_{\odot}$ and a median of 1.17$\times 10^{10}~M_{\odot}$. The stellar mass distribution of the HyMoRS sample, shown in Figure \ref{fig:mstar}, is consistent with that of massive elliptical galaxies. Radio galaxies with elliptical hosts typically exhibit stellar masses ranging from approximately \(10^9\) to nearly \(10^{12}~M_{\odot}\) \citep{Best2006, Sabater2019}. 

\subsubsection{Star Formation Rates}
\label{subsubsec:SFR}
Star formation rate (SFR) is a crucial parameter for understanding galaxy evolution, as it determines the rate of stellar mass buildup and how gas accretion, mergers, and feedback processes shape its growth. The MIR emission, particularly in the W3 (12 $\mu$m) and W4 (22 $\mu$m) bands, effectively traces star formation as it captures the thermal emission from dust heated by young, massive stars. Since HyMoRS galaxies often host AGN, it is crucial to mitigate AGN contamination in the infrared emission to obtain an accurate SFR estimate. high-excitation radio galaxies (HERGs) in particular exhibit strong non-thermal contributions in the MIR bands, which can lead to an overestimation of SFR if not accounted for. Therefore, we identify and remove HERGs using WISE color-color selection (W1$-$W2 vs. W2$-$W3) to distinguish AGN-dominated sources (i.e., W1$-$W2 $>$ 0.5) from star-forming galaxies. After HERGs are removed, we estimate the SFR using the MIR luminosity calibration. We adopt the relation from \citet{Jarrett13} to estimate SFR, which provides an empirical conversion between the 12 $\mu$m luminosity and the SFR previously used for radio galaxies \citep[e.g.,][]{nair24}:

\begin{equation}
\mathrm{SFR}_{\mathrm{W3}} \ (\mathrm{M}_\odot \ \mathrm{yr}^{-1}) = 4.91 \times 10^{-10} \, \nu L_{W3} \ (\mathrm{L}_\odot)
\end{equation}

Using the stellar mass and SFR estimates, we also computed the specific star formation rate (sSFR), which is defined as:  

\begin{equation}
    \text{sSFR} = \frac{\text{SFR}}{M_*} \quad [\text{yr}^{-1}]
    \label{equ:ssfr}
\end{equation}
where \(\text{SFR}\) is the star formation rate in \(M_{\odot} \, \text{yr}^{-1}\) and \(M_*\) is the stellar mass in \(M_{\odot}\). The sSFR measures the star formation activity relative to the stellar mass of a galaxy, provides a better tracer of the cold gas fraction than the global SFR, and thus serves as a robust proxy for short-term gas availability and accretion mode.

We estimated SFRs and sSFRs for the HyMoRS sample presented in this study, as well as for the sources from \citet{Pa22} and \citet{Ka17}. The SFR and sSFR estimates are listed in Columns~13 and 14 of Table~\ref{tab:ir_host}, respectively. The SFR of the combined HyMoRS sample ranges from \(~0.2\) to \(55 \, M_{\odot} \, \text{yr}^{-1}\), with a mean of \(11 \, M_{\odot} \, \text{yr}^{-1}\) and a median of \(5.45 \, M_{\odot} \, \text{yr}^{-1}\). The sSFR for the combined HyMoRS sample varies from \(1.78 \times 10^{-11}\) to \(1.77 \times 10^{-9} \, \text{yr}^{-1}\), with a mean value of \(5.77 \times 10^{-10} \, \text{yr}^{-1}\) and a median of \(3.92 \times 10^{-10} \, \text{yr}^{-1}\). Our result is consistent with \citet{Mingo2022}, who reported substantial SFRs (2.5-58 M$_\odot$ yr$^{-1}$) in typical star-forming radio galaxies, and with \citet{Stroe22}, whose spectroscopic observations likewise showed that HyMoRS display elevated star formation activity, reinforcing their association with star-forming galaxies. Moreover, \citet{Mingo2022} suggested that HERGs with high specific star formation rates (log(sSFR) $> -10~\mathrm{yr}^{-1}$) are preferentially found in star-forming galaxies, regardless of their radio morphology, which is consistent with previous findings\citep[e.g.][]{Janssen2012,gurkan2014,Mingo2016,Williams2018}. Of the subset with W1$-$W2 $< 0.5$ in the combined HyMoRS sample, we found that 52 out of 57 sources (91\%) have log(sSFR) $> -10~\mathrm{yr}^{-1}$. This suggests that the vast majority of these systems are actively star-forming, consistent with the presence of ample cold gas reservoirs capable of fuelling both star formation and AGN activity.

\begin{figure}
\vbox{
\centerline{
\includegraphics[width=9.5cm, origin=c]{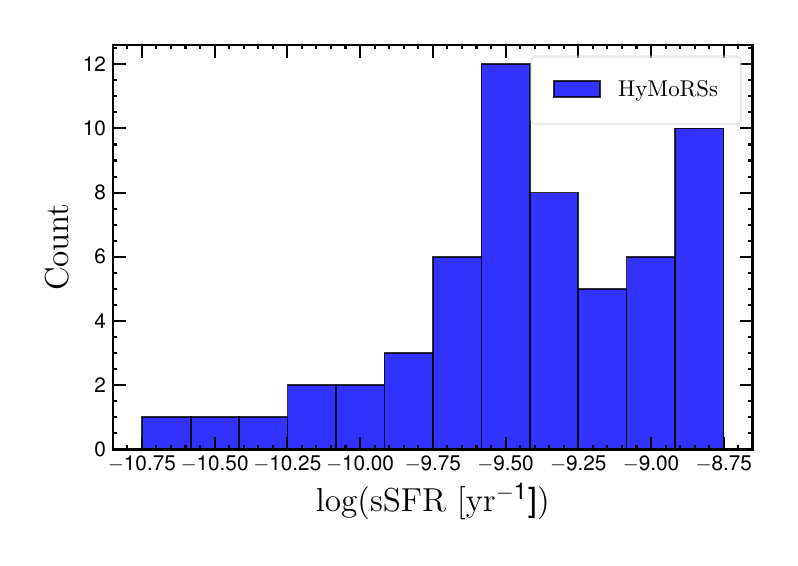}
}
}
\caption{The specific star formation rate (sSFR) distribution of the combined HyMoRS sample includes sources with W1--W2~$\leq$~0.5 from the present work, as well as those reported in \citet{Pa22} and \citet{Ka17}}
\label{fig:ssfr}
\end{figure}

\begin{table*}
    \centering
    %\scriptsize
    \caption{Mid-Infrared host properties of the combined HyMoRS sample include sources from the present work, as well as those reported in \citet{Pa22} and \citet{Ka17}: Col. (1): Serial number. Col. (2): HyMoRS name. Col. (3): redshift of host. Col. (4), (5), (6), (7): Mid-infrared magnitudes at ALLWISE W1, W2, W3, and W4 respectively. Col. (8), (9), (10), (11): corresponding luminosities at ALLWISE bands in log scale. Col. (12): Estimated stellar mass of the host in log scale. Col. (13): Estimated star formation rate of HyMoRSs in log scale. Col. (14): Specific star formation rate of HyMoRSs in log scale. Col. (15): Source reference.}
    \label{tab:ir_host}
    \resizebox{\textwidth}{!}{%
    \begin{tabular}{ccccccccccccccc}
\hline
\hline
No. &      Source &      $z$ &  W1 &  W2 &  W3 &  W4 &    $\log(L_{W1})$ &    $\log(L_{W2})$ &    $\log(L_{W3})$ &    $\log(L_{W4})$ &  $\log(M_{*})$ &  $\log(SFR)$ &   $\log(sSFR)$ &       Ref. \\
~ &      ~ &      ~ &  mag &  mag &  mag &  mag &    ($L_{\odot}$) &    ($L_{\odot}$) &    ($L_{\odot}$) &    ($L_{\odot}$) &  ($M_{\odot}$) &  ($M_{\odot}\text{yr}^{-1}$) &   ($\text{yr}^{-1}$) &       ~ \\
(1) & (2) & (3) &  (4) &  (5) &  (6) &  (7) &    (8) &    (9) &    (10) &    (11) &  (12) &  (13) &   (14) &       (15) \\
\hline
1 & J0016--1208 & 0.988 & 16.32 & 16.08 & 12.56 & 9.09 & 11.06 & 10.76 & 11.05 & 11.56 & 10.55 & 1.74 & --8.82 & This Work \\
2 & J0039--0152 & 0.153 & 15.12 & 14.8 & 12.02 & 8.48 & 9.64 & 9.37 & 9.37 & 9.91 & 8.97 & 0.06 & --8.91 & This Work \\
3 & J0125+1441 & 0.499 & 14.85 & 14.7 & 10.91 & 8.65 & 10.92 & 10.59 & 10.98 & 11.01 & 10.59 & 1.68 & --8.91 & This Work \\
4 & J0153--4837 & 0.323 & 14.94 & 14.73 & 12.51 & 8.78 & 10.44 & 10.13 & 9.9 & 10.52 & 10.0 & 0.59 & --9.41 & This Work \\
5 & J0217--3846 & 0.724 & 15.7 & 15.6 & 12.64 & 8.98 & 10.97 & 10.62 & 10.69 & 11.27 & 10.74 & 1.38 & --9.36 & This Work \\
6 & J0335--5212 & 0.555 & 15.45 & 15.09 & 12.83 & 9.41 & 10.79 & 10.54 & 10.33 & 10.82 & 10.04 & 1.02 & --9.02 & This Work \\
7 & J0339--2228 & 0.276 & 14.99 & 14.75 & 12.72 & 8.63 & 10.26 & 9.96 & 9.66 & 10.42 & 9.77 & 0.35 & --9.42 & This Work \\
8 & J0400--3616 & 0.604 & 15.65 & 15.46 & 12.46 & 8.79 & 10.8 & 10.48 & 10.57 & 11.16 & 10.38 & 1.26 & --9.12 & This Work \\
9 & J0411--6402 & 0.143 & 12.98 & 12.77 & 11.39 & 9.64 & 10.43 & 10.12 & 9.56 & 9.38 & 10.0 & 0.25 & --9.76 & This Work \\
10 & J0509--3621 & 0.427 & 15.55 & 15.24 & 12.63 & 9.22 & 10.48 & 10.21 & 10.14 & 10.63 & 9.85 & 0.83 & --9.02 & This Work \\
11 & J0804+0812 & 0.233 & 14.18 & 13.9 & 12.53 & 8.65 & 10.42 & 10.14 & 9.57 & 10.25 & 9.82 & 0.26 & --9.56 & This Work \\
12 & J0833+0458 & 0.265 & 15.36 & 15.05 & 12.04 & 8.87 & 10.08 & 9.81 & 9.89 & 10.28 & 9.44 & 0.58 & --8.86 & This Work \\
13 & J0835+0436 & 0.761 & 16.44 & 16.47 & 12.4 & 9.03 & 10.73 & 10.32 & 10.83 & 11.31 & 10.77 & 1.52 & --9.24 & This Work \\
14 & J0944--3154 & -- & 14.87 & 14.33 & 12.19 & 8.91 & -- & -- & -- & -- & -- & -- & -- & This Work \\
15 & J0948--3110 & 0.19 & 13.81 & 13.53 & 12.1 & 8.88 & 10.37 & 10.09 & 9.54 & 9.96 & 9.78 & 0.23 & --9.55 & This Work \\
16 & J0953--3253 & -- & 15.32 & 15.44 & 12.37 & 8.48 & -- & -- & -- & -- & -- & -- & -- & This Work \\
17 & J0954--0003 & 0.39 & 14.59 & 14.34 & 11.96 & 8.63 & 10.77 & 10.48 & 10.31 & 10.77 & 10.25 & 1.0 & --9.24 & This Work \\
18 & J1128--0638 & 0.183 & 13.59 & 13.33 & 11.78 & 8.5 & 10.42 & 10.13 & 9.64 & 10.07 & 9.88 & 0.33 & --9.56 & This Work \\
19 & J1243--0100 & 1.336 & 16.51 & 15.77 & 12.36 & 8.74 & 11.3 & 11.21 & 11.45 & 12.03 & -- & -- & -- & This Work \\
20 & J1313--2755 & 0.197 & 13.92 & 13.71 & 12.15 & 8.79 & 10.36 & 10.05 & 9.56 & 10.03 & 9.92 & 0.25 & --9.67 & This Work \\
21 & J1338--2926 & -- & 14.84 & 14.83 & 12.14 & 8.98 & -- & -- & -- & -- & -- & -- & -- & This Work \\
22 & J1348--2409 & 0.5 & 15.39 & 15.09 & 11.96 & 8.75 & 10.71 & 10.43 & 10.57 & 10.98 & 10.09 & 1.26 & --8.82 & This Work \\
23 & J1445--1135 & -- & 15.26 & 15.22 & 12.73 & 9.09 & -- & -- & -- & -- & -- & -- & -- & This Work \\
24 & J1606--1844 & 0.286 & 14.01 & 13.79 & 11.73 & 8.56 & 10.69 & 10.38 & 10.09 & 10.48 & 10.22 & 0.78 & --9.44 & This Work \\
25 & J1626--3633 & 0.043 & 11.51 & 11.54 & 11.26 & 8.75 & 9.92 & 9.52 & 8.51 & 8.64 & 9.95 & --0.8 & --10.75 & This Work \\
26 & J1959--1852 & 0.61 & 15.57 & 14.69 & 11.98 & 8.62 & 10.84 & 10.8 & 10.77 & 11.23 & -- & -- & -- & This Work \\
27 & J2130--2801 & -- & 15.42 & 15.35 & 12.45 & 8.85 & -- & -- & -- & -- & -- & -- & -- & This Work \\
28 & J2137--5405 & 0.404 & 14.94 & 14.76 & 12.67 & 8.94 & 10.67 & 10.34 & 10.07 & 10.68 & 10.29 & 0.76 & --9.53 & This Work \\
29 & J2223--4543 & 0.369 & 14.69 & 14.54 & 11.94 & 8.75 & 10.68 & 10.34 & 10.26 & 10.66 & 10.35 & 0.95 & --9.4 & This Work \\
30 & J2226+1310 & 0.214 & 14.87 & 14.56 & 12.27 & 9.1 & 10.06 & 9.79 & 9.59 & 9.98 & 9.42 & 0.28 & --9.14 & This Work \\
31 & J2231--1406 & 0.407 & 15.21 & 15.06 & 12.56 & 8.14 & 10.57 & 10.23 & 10.12 & 11.01 & 10.24 & 0.81 & --9.43 & This Work \\
32 & J2301--5334 & 0.24 & 14.14 & 13.83 & 12.31 & 9.07 & 10.46 & 10.19 & 9.69 & 10.11 & 9.83 & 0.38 & --9.45 & This Work \\
33 & J2319--4632 & 0.877 & 15.26 & 15.23 & 12.35 & 9.1 & 11.35 & 10.97 & 11.01 & 11.43 & 11.27 & 1.7 & --9.57 & This Work \\
34 & J2336--3735 & -- & 16.18 & 15.91 & 12.27 & 9.0 & -- & -- & -- & -- & -- & -- & -- & This Work \\
35 & J0201+0833 & -- & 12.75 & 12.74 & 12.07 & 9.02 & -- & -- & -- & -- & -- & -- & -- & \citet{Pa22} \\
36 & J0257+0638 & -- & 13.32 & 13.03 & 11.58 & 8.07 & -- & -- & -- & -- & -- & -- & -- & \citet{Pa22} \\
37 & J0715+5528 & -- & 15.14 & 14.36 & 11.89 & 8.2 & -- & -- & -- & -- & -- & -- & -- & \citet{Pa22} \\
38 & J0756+3901 & 0.45 & 14.85 & 14.95 & 12.15 & 8.79 & 10.81 & 10.38 & 10.39 & 10.85 & 10.96 & 1.08 & --9.89 & \citet{Pa22} \\
39 & J0838+3253 & 0.21 & 13.62 & 13.33 & 12.35 & 8.85 & 10.54 & 10.26 & 9.54 & 10.06 & 9.94 & 0.23 & --9.71 & \citet{Pa22} \\
40 & J0855+4911 & 0.09 & 12.36 & 12.28 & 11.36 & 9.21 & 10.25 & 9.88 & 9.14 & 9.12 & 10.07 & --0.17 & --10.24 & \citet{Pa22} \\
41 & J0914+1006 & 0.31 & 14.06 & 13.63 & 11.81 & 8.81 & 10.75 & 10.53 & 10.14 & 10.46 & 9.88 & 0.83 & --9.04 & \citet{Pa22} \\
42 & J1000+3959 & -- & 16.27 & 15.63 & 11.55 & 8.31 & -- & -- & -- & -- & -- & -- & -- & \citet{Pa22} \\
43 & J1027+1033 & 0.11 & 13.44 & 13.37 & 11.73 & 7.57 & 10.0 & 9.63 & 9.17 & 9.96 & 9.83 & --0.14 & --9.97 & \citet{Pa22} \\
44 & J1029+2954 & 0.46 & 15.25 & 14.98 & 12.06 & 9.07 & 10.68 & 10.39 & 10.44 & 10.76 & 10.12 & 1.13 & --8.98 & \citet{Pa22} \\
45 & J1106+1355 & 0.12 & 12.21 & 12.1 & 11.58 & 9.1 & 10.57 & 10.22 & 9.32 & 9.43 & 10.33 & 0.01 & --10.32 & \citet{Pa22} \\
46 & J1156+3910 & 0.42 & 14.77 & 14.59 & 12.67 & 9.15 & 10.77 & 10.45 & 10.1 & 10.64 & 10.38 & 0.8 & --9.58 & \citet{Pa22} \\
47 & J1229+3137 & 0.5 & 15.21 & 15.05 & 12.16 & 8.36 & 10.78 & 10.45 & 10.49 & 11.13 & 10.43 & 1.18 & --9.25 & \citet{Pa22} \\
48 & J1236+5524 & 0.32 & 14.44 & 14.21 & 12.45 & 9.36 & 10.63 & 10.33 & 9.91 & 10.28 & 10.17 & 0.61 & --9.56 & \citet{Pa22} \\
49 & J1249+0932 & 0.23 & 14.17 & 13.92 & 12.11 & 8.89 & 10.41 & 10.12 & 9.72 & 10.14 & 9.88 & 0.41 & --9.47 & \citet{Pa22} \\
50 & J1336+2329 & 0.61 & 15.12 & 14.71 & 12.07 & 8.72 & 11.02 & 10.79 & 10.73 & 11.2 & 10.17 & 1.42 & --8.75 & \citet{Pa22} \\
51 & J1435+5508 & 0.14 & 12.76 & 12.63 & 11.81 & 9.01 & 10.5 & 10.16 & 9.37 & 9.61 & 10.21 & 0.06 & --10.15 & \citet{Pa22} \\
52 & J1538+2144 & 0.47 & 15.21 & 14.99 & 12.7 & 9.53 & 10.71 & 10.41 & 10.21 & 10.6 & 10.24 & 0.9 & --9.34 & \citet{Pa22} \\
53 & J1541+4327 & 0.27 & 14.17 & 13.89 & 12.73 & 8.97 & 10.57 & 10.29 & 9.64 & 10.26 & 10.0 & 0.33 & --9.67 & \citet{Pa22} \\
54 & J1638+3753 & 0.16 & 13.17 & 12.96 & 12.03 & 8.94 & 10.46 & 10.15 & 9.4 & 9.77 & 10.02 & 0.1 & --9.92 & \citet{Pa22} \\
55 & J1657+4319 & 0.2 & 13.95 & 13.7 & 12.56 & 8.99 & 10.36 & 10.07 & 9.41 & 9.96 & 9.83 & 0.1 & --9.73 & \citet{Pa22} \\
56 & J1715+2751 & -- & 16.34 & 16.24 & 12.49 & 8.96 & -- & -- & -- & -- & -- & -- & -- & \citet{Pa22} \\
57 & J2147+1135 & 0.36 & 15.14 & 14.82 & 11.96 & 8.84 & 10.47 & 10.2 & 10.23 & 10.6 & 9.82 & 0.92 & --8.9 & \citet{Pa22} \\
58 & J1136--0328 & 0.82 & 16.04 & 16.17 & 12.06 & 8.41 & 10.97 & 10.52 & 11.05 & 11.63 & 11.21 & 1.74 & --9.47 & \citet{Pa22} \\
59 & J1234--0804 & -- & 13.89 & 13.61 & 12.01 & 8.59 & -- & -- & -- & -- & -- & -- & -- & \citet{Pa22} \\
60 & J1600--0144 & -- & 15.53 & 15.22 & 12.01 & 8.32 & -- & -- & -- & -- & -- & -- & -- & \citet{Pa22} \\
61 & J2129--0549 & -- & 14.17 & 13.92 & 12.46 & 8.96 & -- & -- & -- & -- & -- & -- & -- & \citet{Pa22} \\
62 & J2308--0527 & 0.09 & 12.84 & 12.77 & 12.39 & 8.79 & 10.05 & 9.69 & 8.72 & 9.29 & 9.88 & --0.58 & --10.46 & \citet{Pa22} \\
63 & J0238+0233 & 0.209 & 12.22 & 11.24 & 8.55 & 5.82 & 11.1 & 11.1 & 11.06 & 11.27 & -- & -- & -- & \citet{Ka17} \\
64 & J0724+3803 & 0.241 & 13.76 & 13.07 & 10.61 & 7.53 & 10.62 & 10.5 & 10.37 & 10.73 & -- & -- & -- & \citet{Ka17} \\
65 & J0822+5311 & 0.138 & 14.45 & 14.06 & 11.97 & 8.77 & 9.81 & 9.57 & 9.29 & 9.69 & 9.03 & --0.02 & --9.05 & \citet{Ka17} \\
\hline
\end{tabular}}
\end{table*}

\setcounter{table}{1}
\begin{table*}[t]
    \centering
    %\scriptsize
    \caption{(Continued.)}
    \label{tab:ir_host}
    \resizebox{\textwidth}{!}{%
    \begin{tabular}{ccccccccccccccc}
\hline
\hline
No. &      Source &      $z$ &  W1 &  W2 &  W3 &  W4 &    $\log(L_{W1})$ &    $\log(L_{W2})$ &    $\log(L_{W3})$ &    $\log(L_{W4})$ &  $\log(M_{*})$ &  $\log(SFR)$ &   $\log(sSFR)$ &       Ref. \\
~ &      ~ &      ~ &  mag &  mag &  mag &  mag &    ($L_{\odot}$) &    ($L_{\odot}$) &    ($L_{\odot}$) &    ($L_{\odot}$) &  ($M_{\odot}$) &  ($M_{\odot}\text{yr}^{-1}$) &   ($\text{yr}^{-1}$) &       ~ \\
(1) & (2) & (3) &  (4) &  (5) &  (6) &  (7) &    (8) &    (9) &    (10) &    (11) &  (12) &  (13) &   (14) &       (15) \\
\hline
66 & J0833+0458 & 0.227 & 15.36 & 15.05 & 12.04 & 8.87 & 9.92 & 9.65 & 9.74 & 10.13 & 9.29 & 0.43 & --8.86 & \citet{Ka17} \\
67 & J0847+1831 & -- & 16.91 & 16.09 & 11.76 & 7.84 & -- & -- & -- & -- & -- & -- & -- & \citet{Ka17} \\
68 & J0859+2927 & 0.272 & 15.06 & 14.65 & 12.3 & 8.5 & 10.22 & 9.99 & 9.82 & 10.46 & 9.39 & 0.51 & --8.88 & \citet{Ka17} \\
69 & J0914+5229 & 0.607 & 15.31 & 15.02 & 12.14 & 8.89 & 10.94 & 10.66 & 10.7 & 11.12 & 10.36 & 1.39 & --8.97 & \citet{Ka17} \\
70 & J0942+0627 & 0.359 & 14.83 & 14.59 & 12.42 & 8.68 & 10.59 & 10.29 & 10.05 & 10.66 & 10.08 & 0.74 & --9.35 & \citet{Ka17} \\
71 & J1034+2518 & 0.395 & 15.13 & 14.17 & 10.93 & 7.71 & 10.57 & 10.56 & 10.74 & 11.15 & -- & -- & -- & \citet{Ka17} \\
72 & J1055+3726 & 0.589 & 14.65 & 13.57 & 10.64 & 8.7 & 11.17 & 11.21 & 11.27 & 11.17 & -- & -- & -- & \citet{Ka17} \\
73 & J1058+2445 & 0.201 & 14.38 & 13.67 & 11.63 & 8.1 & 10.19 & 10.09 & 9.79 & 10.32 & -- & -- & -- & \citet{Ka17} \\
74 & J1203+2343 & 0.177 & 13.78 & 13.47 & 11.78 & 8.49 & 10.31 & 10.04 & 9.6 & 10.04 & 9.67 & 0.29 & --9.38 & \citet{Ka17} \\
75 & J1224+0203 & 0.452 & 14.13 & 13.94 & 12.36 & 8.47 & 11.11 & 10.79 & 10.3 & 10.98 & 10.71 & 0.99 & --9.72 & \citet{Ka17} \\
76 & J1226+0429 & 0.517 & 13.06 & 11.9 & 9.03 & 6.57 & 11.68 & 11.75 & 11.78 & 11.88 & -- & -- & -- & \citet{Ka17} \\
77 & J1233+0603 & 0.269 & 15.68 & 14.94 & 11.1 & 7.75 & 9.96 & 9.87 & 10.28 & 10.74 & -- & -- & -- & \citet{Ka17} \\
78 & J1234+2222 & 0.881 & 15.85 & 15.71 & 12.69 & 8.72 & 11.12 & 10.78 & 10.87 & 11.59 & 10.81 & 1.57 & --9.24 & \citet{Ka17} \\
79 & J1314+0204 & 0.982 & 16.11 & 15.38 & 12.41 & 8.82 & 11.13 & 11.03 & 11.1 & 11.66 & -- & -- & -- & \citet{Ka17} \\
80 & J1443+1440 & 0.425 & 16.33 & 16.58 & 12.67 & 9.2 & 10.16 & 9.67 & 10.12 & 10.63 & 10.62 & 0.81 & --9.81 & \citet{Ka17} \\
81 & J1449+5019 & 0.329 & 14.31 & 13.58 & 11.3 & 8.93 & 10.71 & 10.61 & 10.41 & 10.48 & -- & -- & -- & \citet{Ka17} \\
82 & J1504+5649 & 0.359 & 12.96 & 11.93 & 9.19 & 6.94 & 11.34 & 11.36 & 11.34 & 11.36 & -- & -- & -- & \citet{Ka17} \\
83 & J1511+3355 & 0.623 & 14.9 & 14.47 & 11.97 & 9.08 & 11.13 & 10.91 & 10.8 & 11.08 & 10.25 & 1.49 & --8.77 & \citet{Ka17} \\
84 & J1527+1822 & 0.445 & 15.33 & 14.7 & 12.14 & 8.76 & 10.61 & 10.47 & 10.38 & 10.85 & -- & -- & -- & \citet{Ka17} \\
85 & J1534+3334 & 0.21 & 14.7 & 14.34 & 12.75 & 8.93 & 10.11 & 9.86 & 9.38 & 10.03 & 9.37 & 0.07 & --9.3 & \citet{Ka17} \\
\hline
\end{tabular}}
\end{table*}

\subsection{Notes on HyMoRS with The Signature of Multi-peak and Distorted Morphology}
\label{sec:restarted}
In the following, we discuss some HyMoRS with signatures of multi-peak emission and distorted morphology.
Radio galaxies with multi-peak emission sometimes refer to recurrent activity in the source or it may be produced when the jet of radio galaxy interacts with surrounding media. Recurrent or episodic activity periods are intense activity in the central AGN followed by quiescent phases. This cyclical behavior results in observable changes in the radio emissions and morphology of the galaxy over time. 

\subsubsection{J0125+1441}
The north-western side lobe of J0125+1441 displays a compact hotspot at its far end, i.e., at the FR II lobe. The south-eastern side structure is bent ($\sim$20$^{\circ}$) with the brightest peak close to the center i.e., the FR I lobe. Apart from the brightest peak close to the center, the south-eastern side lobe also displays a secondary peak close to the end of the structure. The photometric redshift of the source is 0.499, with a linear size of 0.54 Mpc. Using the MAL survey, the spectral index between 1000 MHz and 1400 MHz is measured as 0.61. 

\subsubsection{J0217--3846}
The western side lobe of J0217--3846 displays a compact hotspot component at the outer edge as seen in the FR II lobe. Similar to J0125+1441, the FR I lobe along the north-eastern side displays a brightest peak close to the radio core and a secondary peak close to the edge of the FR I lobe. The radio core is detected for this source. The FR I lobe is distorted from the position of the radio core. The available photometric redshift of the source is 0.724 \citep{Du22} with a linear size of 433 kpc. Using the MAL survey, the spectral index between 1000 MHz and 1400 MHz is measured as 1.01. 

\subsubsection{J1313--2755}
The south-western side lobe of J1313--2755 displays the brightest compact hotspot i.e., the FR II lobe at its outer edge along with a radio peak close to the core. The north-eastern side lobe displays the brightest peak close to the core i.e., the FR I lobe. The FR I lobe also displays a faint secondary peak at the far end of the structure. The radio core is not detected for this source. No redshift is available for this source. The total angular extent of the source is 114.5 arcsec. Using the MAL survey, the spectral index between 1000 MHz and 1400 MHz is measured as 0.85.
    
\subsubsection{J1338--2927}
The northern side lobe of J1338--2927 is an FR II lobe with a compact hotspot at its outer edge. The south-western side lobe is an FR I lobe that displays a tail-like structure with the brightest peak in the inner part mixed with the radio core and a secondary radio peak at the far end of the structure. The FR I lobe is bent with $\sim20^{\circ}$ bending angle. No redshift information is available for this source. The total angular extent of the source is 66.6 arcsec. The spectral index between 1000 MHz and 1400 MHz using the MAL survey is measured as 1.38.

\subsubsection{J2226+1310}
The eastern side lobe of J2226+1310 displays the brightest compact hotspot at its outer edge, indicating that the lobe is of FR II type. The brightest peak in the south-western side lobe is close to the core indicating the lobe to be of FR I type. The FR I lobe also displays a faint secondary peak at the far end of the structure. A bright compact radio core is detected between the FR II and FR I lobes. The FR I and FR II lobes of the source seem to be distorted with $\sim10^{\circ}$ bending angle from the main axis of propagation with respect to the radio core. The available photometric redshift of the source is $z=0.214$ with a linear size of 247 kpc. The source is detected only at 1006 MHz and not at 1348 MHz. So, the spectral index between 1000 MHz and 1400 MHz is not available for this source. 
\subsubsection{J2231--1406}
The FR II lobe i.e., the brightest compact hotspot at the outer edge along the south-eastern side and the FR I lobe along the north-western side with the brightest peak close to the core can be seen for J2231--1406. At 1348 MHz of the MAL survey, a radio peak close to the core is detected along the FR II lobe and a faint secondary radio peak is detected at the outer edge along the FR I lobe. A bright compact radio core is detected between the FR I and FR II lobes. The FR I lobe is bent with $\sim20^{\circ}$ bending angle from the main axis of propagation with respect to the radio core. The available photometric redshift of the source is $z=0.407$ with a linear size of the source of 360 kpc. The spectral index between 1000 MHz and 1400 MHz for this source is 0.6.

\subsubsection{J2336--3736}
The MAL survey at 1006 MHz detects the brightest compact hotspot at its outer edge, i.e., the FR II lobe in J2336--3736, along the north-western side, whereas the FR I lobe is located along the south-eastern side. The FR I lobe contains a long tail-like feature with multiple radio peaks with bimodal intensity distribution from the core to the outer edge of the structure. The multiple radio peaks are detected at both frequencies of the MAL survey, i.e., at 1006 MHz and 1348 MHz. A bright compact radio core is detected between the FR II and FR I lobes. The source is bent with $\sim10^{\circ}$ bending angle with respect to the radio core. No redshift information is available for this source. The total angular extent of the source is 78 arcsec. Using the MAL survey, the spectral index between 1000 MHz and 1400 MHz is measured as 0.71.

\begin{figure}
\vbox{
\centerline{
\includegraphics[width=9.5cm, origin=c]{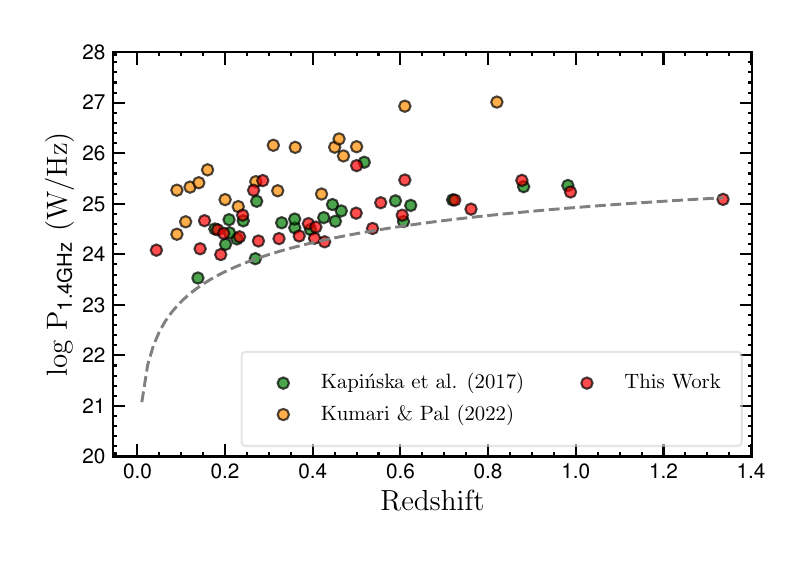}
}
}
\caption{Distribution of radio power with the corresponding redshift of the sources present in the current paper along with \citet{Pa22} and \citet{Ka17}. The grey dashed line represents the MALS limit of detecting HyMoRSs with low surface brightness.} 
\label{fig:P-z}
\end{figure}

\begin{table*}
    \centering
    \scriptsize
    \caption{Mals HyMoRSs in galaxy clusters: Col. (1): Serial number. Col. (2): HyMoRS name. Col. (3): Cluster name as per \citet{wh24} Col. (4) and Col. (5): RA and DEC of cluster position. Col. (6): Redshift of galaxy cluster. Col. (7): Stellar mass of BCG. Col. (8): Cluster radius. Col. (9): Cluster richness. Col. (10): Cluster mass. Col. (11): Number of member galaxy candidates within $r_{500}$. Col. (12): Comoving number density of galaxies within $r_{500}$.}
    \label{tab:cluster}
    \begin{tabular}{cccccccccccc}
\hline
\hline
No. &      Source &            Cluster name &   RA &  DEC &     $z_{Cl}$ &  log$M_{*}$ &  $r_{500}$ &  $\lambda_{500}$ &  M$_{500}$ &  N$_{500}$ &    $\rho$ \\
~ &  ~ & ~ &   ($^\circ$) &  ($^\circ$) &     ~ &  ($M_{\odot}$) &  (Mpc) &  ~ &  ($10^{14}~M_{\odot}$) &  ~ &    (Mpc$^{-3}$) \\
(1) &      (2) &  (3) &   (4) &  (5) &     (6) &  (7) &  (8) &  (9) &  (10) &  (11) &    (12) \\
\hline
1 &  J0039--0152 &  J003922.8--015213 &    09.84510 &  --01.87016 &  0.1537 &  11.63 &  0.90 &   51.01 &  2.27 &    36 &  11.67 \\
2 &  J0217--3846 &  J021759.3--384640 &   34.49702 & --38.77769 &  0.6995 &  11.47 &  0.47 &   10.81 &  0.50 &     ~7 &  15.99 \\
3 &  J0411--6402 &  J041104.2--640241 &   62.76751 & --64.04463 &  0.1601 &  11.72 &  0.84 &   33.75 &  1.52 &    32 &  13.08 \\
4 &  J0509--3621 &  J050925.6--362107 &   77.35662 & --36.35197 &  0.4372 &  11.17 &  0.47 &   11.37 &  0.53 &     ~9 &  21.37 \\
5 &  J0804+0812 &  J080448.0+081254 &  121.20019 &   ~08.21501 &  0.2325 &  11.73 &  0.57 &   18.04 &  0.83 &     ~8 &  10.42 \\
6 &  J1128--0638 &  J112853.5--063852 &  172.22305 &  --06.64790 &  0.1993 &  11.72 &  0.57 &   17.51 &  0.81 &     ~6 &   07.61 \\
7 &  J1445--1135 &  J144512.0--113545 &  221.29990 & --11.59570 &  0.6931 &  11.55 &  0.59 &   26.89 &  1.22 &    13 &  15.27 \\
8 &  J2226+1310 &  J222610.8+131026 &  336.54520 &  ~13.17394 &  0.2268 &  11.26 &  0.50 &   12.56 &  0.58 &    10 &  19.10 \\
9 &  J2319--4632 &  J231942.1--463301 &  349.92551 & --46.55018 &  0.8112 &  11.65 &  0.46 &   17.11 &  0.79 &     ~7 &  17.06 \\
\hline
\end{tabular}
\end{table*}

\subsection{HyMoRS in the Cluster Environment}
\label{subsec:cluster}
HyMoRSs are believed to be shaped not only by the intrinsic properties of their central AGN but also by the influence of large-scale environments. In particular, asymmetries in the ambient medium, such as those found in galaxy clusters, may contribute to the morphological dichotomy observed between their radio lobes.

To explore this possibility, we cross-matched our sample of HyMoRS with the updated galaxy cluster catalog from \citet{wh24}. We find that nine out of thirty-six sources (\( \sim25\% \)) are located near the centers of galaxy clusters and are hosted by Brightest Cluster Galaxies (BCGs), including one system (J2319--4632) with giant radio jets extending up to 811~kpc. The subset of our HyMoRS sample associated with BCGs, along with their cluster properties, is summarized in Table~\ref{tab:cluster}. M$_{500}$ in Table~\ref{tab:cluster} indicates the mass of a galaxy cluster within a radius of R$_{500}$, where the density is 500 times the critical density of the Universe. The cluster masses (M$_{500}$) for these sources range from \( 0.5 \times 10^{14} \, M_\odot \) to \( 2.27 \times 10^{14} \, M_\odot \), with one source (J0039--0152) residing in a particularly massive cluster that exceeds the commonly used threshold limit of \( 2 \times 10^{14} \, M_\odot \). We also estimated the comoving number density of HyMoRS located in cluster environments, which ranges from \( 7.61 \) to \( 21.37 \, \mathrm{Mpc}^{-3} \), indicating their presence in significantly dense regions. Moreover, we find that sixteen HyMoRSs out of the thirty-six in our sample exhibit bending from their primary axis of propagation. Fourteen of these sources show bending angles between \( 10^\circ \) and \( 20^\circ \), while two (J0954--0003 and J2357--0725) exhibits larger bending angle of \( 52^\circ \) and \( 60^\circ \), respectively. These deflections likely indicate interactions with dense or cluster-like asymmetric environments.

\section{Discussions}
\label{sec:discuss}
The current article reports thirty-six hybrid morphology radio sources using the MAL survey. This is the largest sample of HyMoRS discovered in the southern sky. In this section, we discuss and interpret the physical properties observed in these HyMoRSs. In Section \ref{subsec:raddiscuss}, we discuss the observed radio properties; in Section \ref{subsec:hostprop}, we examine the properties of the host galaxies; and in Section \ref{subsec:formationscenarios}, we explore the possible formation scenarios of HyMoRSs.

\begin{figure}
\vbox{
\centerline{
\includegraphics[width=9.5cm, origin=c]{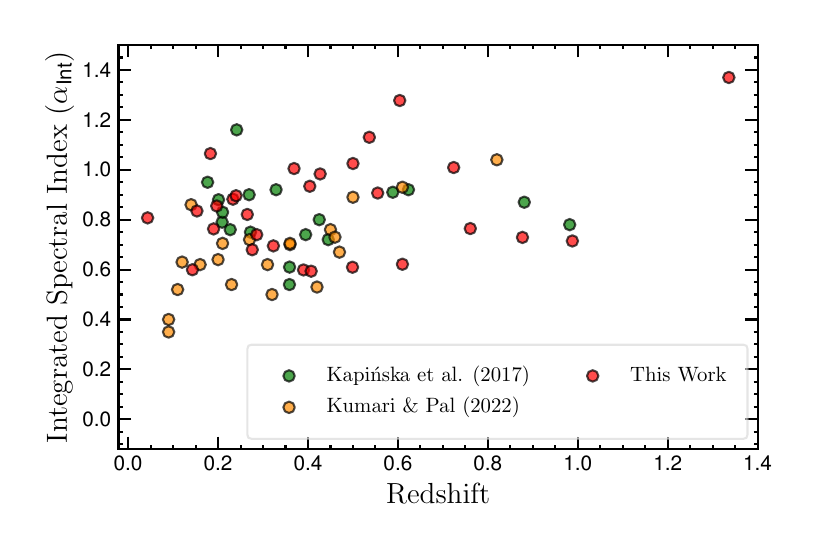}
}
}
\caption{Variation of the integrated spectral index with redshift for the HyMoRSs in the present paper along with \citet{Pa22} and \citet{Ka17}.}
\label{fig:z_alpha}
\end{figure}

\subsection{Interpretation of Radio Properties}
\label{subsec:raddiscuss}
Figure \ref{fig:P-z} displays the distribution of HyMoRS in the $\log~\mathrm{P_{1.4GHz}}-z$ plane. Most of the HyMoRS lie within the redshift range of 0.1 to 0.6, with a decrease in the number of sources as the redshift increases beyond $z = 0.6$. 

Figure~\ref{fig:z_alpha} shows that the integrated spectral index of HyMoRS galaxies increases with redshift, consistent with previous findings for powerful radio sources \citep{Lai80, Bl99}. This correlation suggests that higher-redshift sources tend to exhibit steeper spectra, likely due to enhanced energy losses affecting relativistic electrons. During injection, electrons undergo first-order Fermi acceleration at both relativistic and non-relativistic shock fronts, leading to a power-law energy distribution \citep{At98}. As redshift increases, inverse Compton (IC) losses become more significant due to the increased energy density of the cosmic microwave background (CMB), which scales as $(1+z)^4$. When high-energy electrons in the plasma interact with CMB photons, they lose energy more efficiently through inverse-Compton scattering, which causes the radio spectra to steepen \citep{At98, Mora2018}. Additionally, powerful radio galaxies exhibit stronger magnetic fields, which enhance synchrotron losses. Furthermore, the presence of denser gas clumps at high redshift slows jet propagation, causing hotspots and lobes to remain in a high-pressure environment for extended periods, further steepening the electron energy spectrum \citep{ks87, At98, Go12}.

\subsection{Properties of the Host Galaxies of HyMoRS}
\label{subsec:hostprop}
Here, we discuss the host properties of the combined sample of HyMoRS, which includes sources from the present study as well as those from \citet{Pa22} and \citet{Ka17}. In Section~\ref{subsubsec:hostprop}, we presented insights into the accretion modes of HyMoRS hosts, and in Section~\ref{subsubsec:ssfrevo}, we examined the evolution of specific star formation rate (sSFR) with redshift.

\begin{figure}
\vbox{
\centerline{
\includegraphics[width=9.5cm, origin=c]{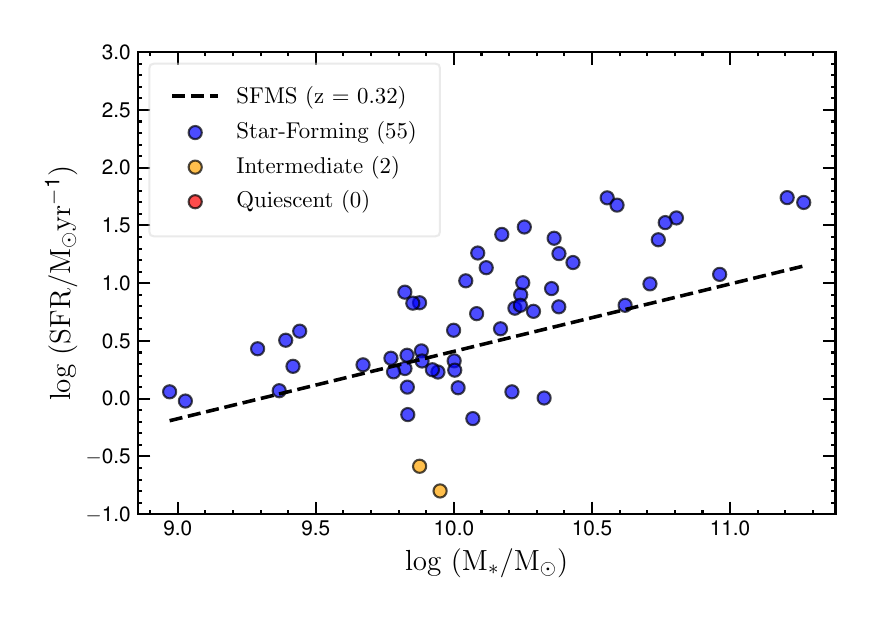}
}
}
\caption{The distribution of star formation rates (SFRs) as a function of stellar mass is shown for the combined HyMoRS sample, which includes sources with W1--W2~$\leq$~0.5 from the present study, as well as those from \citet{Pa22} and \citet{Ka17}. The black dashed line represents the star-forming main sequence (SFMS), indicating the typical SFR–stellar mass relation for star-forming galaxies, based on the prescription provided by \citet{sp14}.}
\label{fig:sfrmstar}
\end{figure}
\subsubsection{Accretion Modes of Hosts of HyMoRS}
\label{subsubsec:hostprop}
As discussed in Section~\ref{subsubsec:IRcolor}, the majority of the host galaxies of HyMoRSs presented in this study lie in the LERG/SFG region of the WISE color--color plot (see Figure~\ref{fig:WISE_color}). For the combined sample, which includes HyMoRSs from the current work along with those reported by \citet{Pa22} and \citet{Ka17}, we find that 20\% of sources fall in the AGN/HERG region, 23\% in the LERG (ellipticals) region, 48\% in the SFG region, and 8\% in the ULIRG region. This result confirms that most HyMoRS are associated with the SFG region. Furthermore, Figure~\ref{fig:WISE_color} suggests that the majority of HyMoRS classified in the AGN/HERG region are drawn from the sample of \citet{Ka17}. Since we also estimated the SFR and stellar mass for sources with WISE color W1--W2 $<$ 0.5, we further analyzed their accretion modes by plotting them on the SFR–M$_{*}$ plane in Figure \ref{fig:sfrmstar}. The black dashed line represents the star-forming main sequence (SFMS) relation at the median redshift of the HyMoRS sample, as calculated using the relation from \citet{sp14}. We classify star-forming, intermediate, and quiescent galaxies based on their deviation from the SFMS at a given redshift (\(z\)) and stellar mass (\(M_*\)). Galaxies are considered star-forming if their specific star formation rate (sSFR) satisfies \(\log (\rm SFR/SFR_{MS}(t)) \geq -0.4\), while those with \(-1.3 < \log (\rm SFR/SFR_{MS}(t)) \leq -0.4\) are classified as intermediate . Meanwhile, galaxies with \(\log (\rm SFR/SFR_{MS}(t)) < -1.3\) are considered quiescent. The classification based on the SFMS plane reveals that all 52 HyMoRSs are categorized as star-forming and two sources belongs to the intermediate region. Given that we selected sources with W1--W2 $<$ 0.5, effectively filtering out AGN-dominated systems, this result suggests that HyMoRS galaxies are preferentially associated with ongoing star formation. 

The observed star-forming nature of the HyMoRS sample provides insight into the connection between host galaxy properties and accretion modes in AGN. The accretion mode in AGN is primarily influenced by the availability of cold, dense gas in the vicinity of the supermassive black hole. The presence of this cold gas can significantly enhance the accretion rate beyond what is achievable through fueling mechanisms associated solely with the surrounding hot gas medium \citep{Gaspari2013, Gaspari2015, Hardcastle2020}. Additionally, the black hole mass plays a crucial role in AGN than in X-ray binaries, in determining the amount of gas required to transition between radiatively inefficient (RI) and radiatively efficient (RE) accretion modes \citep{Hardcastle2018a}. The sSFR serves as a more reliable indicator of the fractional gas content of a galaxy than the global star formation rate \citep{Abramson2014, Ilbert2015}. Consequently, it can be considered a valuable proxy for assessing the short-term availability of cold gas, which directly impacts the accretion mode. In a recent study, \citet{Mingo2022} suggest that HERGs with high specific star formation rates (log(sSFR) $> -10 ~\text{yr}^{-1}$) predominantly reside in star-forming systems, regardless of their radio morphology. This finding underscores the crucial role of cold gas availability in both star formation and AGN fueling. Moreover, they show that star-forming low-power FR II (FR II-Low) sources predominantly lie near the FR I/II division line ($\rm P_{150} \sim 10^{26}$ W Hz$^{-1}$). As seen in Figure \ref{fig:ssfr}, almost all HyMoRS exhibit log(sSFR) values exceeding $-10~\text{yr}^{-1}$, indicating radiatively efficient accretion and an association with HERG hosts. They exhibit radio luminosities ranging from $10^{24.69}$ W Hz$^{-1}$ to $10^{26.75}$ W Hz$^{-1}$, with a mean value of $10^{25.53}$ W Hz$^{-1}$, which is close to the FR I/II division line. This suggests that HyMoRSs in the present study are likely related to the FR II-Low population in terms of their radio morphology. In summary, the host properties of HyMoRS indicate that they are predominantly HERGs, with star-forming galaxies as hosts and strong Type 1 AGN/quasar features. The result is consistent with the previous finding from spectroscopic data \citep{Stroe22}. 

\begin{figure}
\vbox{
\centerline{
\includegraphics[width=9.5cm, origin=c]{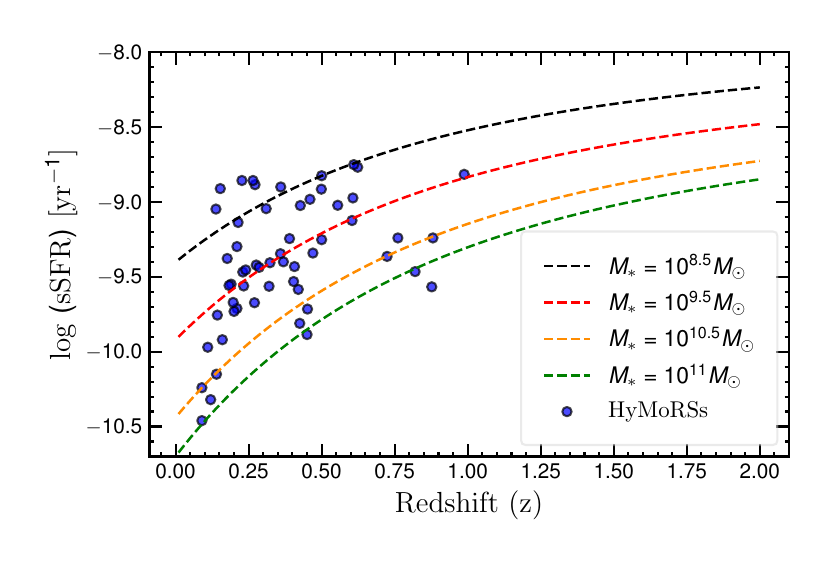}
}
}
\caption{The evolution of the specific star formation rate (sSFR) with redshift is shown for the combined HyMoRS sample, which includes sources with W1--W2~$\leq$~0.5 from the present study, along with those from \citet{Pa22} and \citet{Ka17}. The black dashed lines represent the predicted sSFR evolution of star-forming main sequence (SFMS) galaxies with stellar masses of $10^{8.5}$, $10^{9.5}$, $10^{10.5},$ and $10^{11}\,M_{\odot}$, based on the SFMS normalization calculated using the prescription from \citet{sp14}.}
\label{fig:ssfrz}
\end{figure}

\subsubsection{Evolution of Specific Star Formation Rate}
\label{subsubsec:ssfrevo}
In Figure \ref{fig:ssfrz}, we showed the evolution of sSFR with redshift. We can see a clear decreasing trend of sSFR with cosmic time. To assess the consistency of our sSFR trend with previous studies, we derive sSFR($z$) using the SFMS normalization based on \citet{sp14}. We compute sSFR($z$) for stellar masses of $10^{8.5}$, $10^{9.5}$, $10^{10.5},$ and $10^{11}\,M_{\odot}$, and plotted these as black dashed lines in the figure \ref{fig:ssfrz}. Our results indicate that the evolutionary track of sSFR with redshift in our sample broadly agrees with the predictions based on SFMS normalization.

The decrease in sSFR with cosmic time is primarily driven by the declining gas supply from cosmological accretion that fuels star formation. As redshift increases, the gas accretion rate scales approximately as $\dot{M}_{\rm acc}/M \propto (1+z)^{2.25-2.5}$ \citep{dekel09, dekel13}. However, by $z \lesssim 2$, this rate drops below the level required to sustain the high sSFR observed in many galaxies \citep{dekel09, weinmann11, dave11}. This decline in accretion may be further influenced by the angular momentum of the infalling gas and that of the material returned by the evolved stellar population.

As galaxies evolve, the gas that is accreted tends to have higher angular momentum, causing it to settle primarily in the outer regions of the disk \citep{stewart11}. This spreading out of gas leads to a decrease in the overall gas mass surface density. In simple terms, while the total mass of a galaxy might remain roughly constant, its effective radius grows over time (with an expected scaling of approximately $(1+z)^{-1.5}$) \citep{mo98}, causing the mass, and hence the star formation activity, to be distributed over a larger area. Geometrically, this means that the disk mass surface density (given by the total mass over all epochs divided by the area of the disk) scales approximately as $(1+z)^{3}$. Observations, however, suggest a slightly slower evolution in stellar radii, around $(1+z)^{-1.2\pm0.1}$ \citep[e.g.,][]{mosleh11,mosleh12}. The results suggest that cosmic gas accretion and angular momentum evolution play a fundamental role in regulating star formation in galaxies over cosmic time.
\subsection{Possible Formation Scenarios of HyMoRS}
\label{subsec:formationscenarios}
Here, we explore the possible formation scenarios of HyMoRS. In Section~\ref{subsubsec:orientation1}, we discussed the role of orientation, along with other intrinsic jet properties, in producing the observed asymmetries in HyMoRSs. In Section~\ref{subsubsec:enveffect}, we examined the impact of external environmental influences on the observed radio morphology of HyMoRSs.
\subsubsection{Orientation \& Other Intrinsic Jet Properties}
\label{subsubsec:orientation1}
The observed morphology of radio galaxies is highly dependent on their orientation with respect to the observer. \citet{Ha20} proposed that hybrid morphology may result from a favorable viewing angle, causing one of the FR II jets to appear as an FR I. This orientation also provides a clear view of the narrow and broad-line regions, leading to a Type 1 AGN or quasar optical classification. Recent three-dimensional relativistic hydrodynamic simulations by \citet{bh24} have successfully reproduced the boosted FR I/deboosted FR II hybrid morphology observed in radio galaxies, aligning well with observational trends. As shown in Figures 9(b), (d), and (f) of \citet{bh24}, at an inclination angle of $30^{\circ}$, the receding counterjets appear as edge-brightened structures, while the approaching jets exhibit a BL Lac-like FR I morphology. This result strongly supports the interpretation that HyMoRSs arise due to orientation effects, in addition to other environmental and intrinsic factors.

In Section \ref{subsubsec:orientation}, we estimated the inclination angle ($\phi$) of the HyMoRSs analyzed in this study. Figure \ref{fig:inc} presents the distribution of $\phi$, where we find that all HyMoRSs in our sample have inclination angles greater than $22.15^\circ$. The mean and median inclination angles of our sample are $38^\circ$ and $36^\circ$, respectively. These values align with the theoretical predictions of \citet{bh24} and support the lower limit suggested by \citet{Ce13}. 

HyMoRSs may indeed represent an intrinsically transient phenomenon. It has been suggested that a combination of central engine modulation and the differences in light travel time between the approaching and receding jets of a radio source could influence the observable radio morphology. Three scenarios can be considered to explain the intrinsic nature of HyMoRSs:

(i) The observed asymmetry may arise due to the restarted activity of some radio galaxies, as proposed in the model of \cite{2012AA...544L...2M}. This model suggests that when a radio source is inclined at a moderate angle to our line of sight ($\lesssim45^{\circ}$) and its jet production restarts within $1$--$80\times10^4$~yrs, the resulting difference in light travel time can temporarily create an appearance in which one side exhibits FR II characteristics (representing older emission), while the other shows FR I features (representing newer emission), producing what appears to be a HyMoRS from our perspective. This model has been used to interpret the observed radio structures of at least three well-known HyMoRS sources, J1211+743, 3C249.1, and 3C334, based solely on their radio morphology \cite[e.g.][]{2012AA...544L...2M,2012A&A...545A.132M}. Although the model was originally proposed to question whether the hybrid morphology is the result of different intrinsic AGN behaviours of radio galaxies, it is possible that a fraction of the HyMoRS population may represent a transient phase of AGN activity \cite[see][]{2013A&A...557A..75C}.

(ii) An alternative scenario involves the complete cessation of jet activity \citep{Ka17}, rather than its restart. In this case, if the central engine switches off, the light travel time effects would again lead to an asymmetric appearance: the distant lobe (on the far side) may still retain its FR II-like structure due to the slower fading of hotspots, while the nearer lobe (on the approaching side) may already exhibit a more diffuse, FR I-like morphology as the hotspot fades more rapidly \cite[e.g.][]{2012AA...544L...2M}. The hotspots typically fade on timescales of $\sim10^4$--$10^5$~yrs, whereas the more extended lobe emission can persist for $>10^7$~yrs in the absence of particle re-acceleration.

(iii) A third scenario considers activity amplification \citep{Ka17}, where the AGN undergoes a new phase of enhanced accretion. In such cases, a radio galaxy that was initially of either FR I or FR II type could develop a hybrid appearance depending on the timescales of activity modulation and the light travel time effects. For example, if the source was initially an FR I-type radio galaxy, renewed powerful jet activity could produce an FR II-like morphology on the side closer to the observer, while the far side retains the older FR I structure. Hydrodynamical simulations of supermassive black hole evolution predict highly intermittent accretion rates on a variety of timescales, from bursts within a single accretion episode (lasting $\sim10^6$~yrs) to long-term cycles of activity and quiescence separated by $\sim10^7$--$10^8$~yrs \cite[e.g.][]{2011ApJ...737...26N,2014ApJ...789..150G}. Additionally, \cite{2015MNRAS.451.2517S} suggested that the typical AGN phase, particularly in the optical and X-ray regimes, may last for $\sim10^5$~yrs, which aligns well with the timescales required by the light travel time arguments discussed here.

In this work, we found seven sources (J0125+1441, J0217--3847, J1313--2755, J1338--2927, J2226+1310, J2231--1406, and J2336--3736) among the 36 sources presented possess multiple hotspots along the FR I lobe regions (discussed in Sect. \ref{sec:restarted}). Multiple hotspots along the FR I may arise due to the interaction of the lobe with the dense environment, whereas the other lobe (FR II lobe) grows larger due to the less dense environment. The brightness distribution in the source J2336--3736 is bimodal along the FR I-type lobe: two secondary peaks of emission are located along the edge of the lobe, a decrease in brightness occurs towards the middle of the structure, and the brightest primary acceleration region is located close to the core. In source J1313--2755, an episode of the second AGN activity can be seen in which the second activity, along the FR I, collapses the lobe region of the first activity. In this source, as the FR I lobe was affected by a denser environment, it does not grow larger and may easily interact with the second AGN activity, whereas the other side (FR II) lobe does not interact with the second activity because it grows larger in size. 

Therefore, our result suggests the episodic activity of the AGN may have contributed to the formation of the observed morphology of HyMoRS.  Future high-resolution multi-wavelength follow-up observation is required to study the episodic nature of these sources in detail.

\subsubsection{External Environmental Influence}
\label{subsubsec:enveffect}
The surrounding environment plays a crucial role in shaping the structure of radio galaxies. HyMoRSs serve as key evidence of how external conditions influence their evolution \citep{go02,meliani08}. Previous studies suggest that radio galaxies may initially evolve as FR II-type sources \citep[e.g.,][]{Ka07,Ce13}, with the transition to an FR I morphology occurring due to jet deceleration and disruption \citep[e.g.,][]{li94, meliani08, wang11, perucho12, turner15}. Various mechanisms, including entrainment of surrounding material \citep{wang11}, helical instabilities \citep{perucho12}, and density variations in an inhomogeneous external medium, have been proposed to explain this transition. \citet{meliani08} developed a theoretical model specifically for HyMoRSs, while most deceleration models focus on the general transition from FR II to FR I radio sources. Their study suggests that jet deceleration is caused by a density jump in the external medium. \citet{Ha20} suggested that hybrid morphologies may result from large-scale environmental effects that push back the FR II jets, combined with a favorable orientation of the source. In a comprehensive multi-wavelength analysis of a giant FR II radio galaxy exhibiting HyMoRS-like lobe asymmetry, \citet{seymour2020} found that the FR I–like jet propagates into a dense intracluster medium (ICM) of the neighbouring irregular cluster, likely causing its deceleration. This interaction, combined with the ram pressure exerted by the ICM, distorts the radio jets and produces morphologies that resemble hybrid sources in projection. 

In line with these findings, as discussed in Section \ref{subsec:cluster}, we identified nine HyMoRS candidates that are hosted by BCGs, reinforcing the idea that dense environments play a critical role in shaping hybrid radio morphologies. Although there is currently no comprehensive study for HyMoRSs, the available data presented by \citet{Stroe22} suggest that FR II HyMoRSs tend to reside in relatively dense environments, such as low-mass clusters or galaxy groups. This is supported by the \( M_{500} \) values of our sample, which range from \( 0.5 \times 10^{14} \, M_\odot \) to \( 2.27 \times 10^{14} \, M_\odot \), with only one source exceeding the \( 2 \times 10^{14} \, M_\odot \) threshold commonly used to define massive clusters. 

Moreover, we found that the host galaxies associated with galaxy clusters exhibit high SFRs, ranging from \(1.15\) to \(50 \, M_\odot\,\text{yr}^{-1}\). These high SFRs indicate that the host galaxies still contain substantial gas reservoirs, suggesting that they have not yet undergone the quenching typically associated with dense cluster environments. This scenario is consistent with the idea that these HyMoRS are recent infallers into the cluster, which could explain both their ongoing star formation and their relative motion within the surrounding medium. 

Interestingly, the source J0039--0152, which is associated with the most massive cluster in our sample (\(M_{500} = 2.27 \times 10^{14} \, M_\odot\)), shows the lowest SFR among the cluster-associated HyMoRSs, with a value of \(3.70 \, M_\odot\,\text{yr}^{-1}\). This aligns with the expectation that massive clusters tend to suppress star formation in their member galaxies.

We found that sixteen HyMoRSs in our sample of thirty-six sources exhibit bending away from their main axis of propagation. Among these, fourteen sources show bending angles between \(10^{\circ}\) and \(20^{\circ}\), while the remaining two exhibit more pronounced bending of \(52^{\circ}\) and \(60^{\circ}\) (J0954--0003 and J2357--0725), respectively. Notably, all of these sources bend in a common direction, reminiscent of the morphology observed in wide-angle-tailed (WAT) and narrow-angle-tailed (NAT) radio galaxies. Among the identified bent HyMoRSs, five sources (J1338--2925, J0948--3110, J2301--1405, and J2357--0725) show bending only in the FR I lobe. Of these, four have bending angles of approximately \(10^{\circ}\), while one (J2357--0725) shows a larger bending angle of around \(60^{\circ}\). Such bent jet structures are typically indicative of interaction with a dense surrounding medium \citep{Bh22, Pa23}, suggesting that these HyMoRSs may reside either in galaxy groups or within the intracluster environment, both of which could account for the observed distortions. Future multi-wavelength studies, including X-ray observations of the ICM and deeper radio imaging, are essential to further constrain the role of clusters in shaping HyMoRSs.

\section{Conclusions}
\label{sec:conclusion}
In the present study, we undertook a systematic search for HyMoRSs in the southern sky utilizing data from the MAL survey. This survey was performed using the MeerKAT radio telescope and covered an area of 2289 deg$^{2}$ (1132 deg$^{2}$). We also investigated the host galaxy properties using mid-infrared data from the ALLWISE survey. Moreover, we studied the environmental properties of the HyMoRSs presented in this paper, along with those of previous samples, and discussed different formation scenarios of HyMoRSs. Our key findings from the present work are summarised below:

\begin{itemize}
\item We discovered a sample of 36 HyMoRSs from the MAL survey, representing the largest sample identified in the southern sky. Given that only $\sim$60 HyMoRS were previously known, the present work increases the total known population by $\sim$60\%.

\item The HyMoRSs presented in this work lie within a redshift range of \( z = 0.04 \) to \( 1.34 \), with total radio powers ranging from \(9.9 \times 10^{23}\) to \(5.7 \times 10^{25}~\, \mathrm{W\,Hz^{-1}}\), and integrated spectral indices between \( 0.59 \) and \( 1.39 \).

\item The HyMoRSs presented in this study exhibit projected linear sizes ranging from 105 to 811~kpc. Notably, one source (J2319--4632) hosts large-scale radio jets, extending up to 811~kpc, thereby qualifying to be classified as a giant radio galaxy.

\item We conducted a statistical comparison of the spectral indices between the two lobes of HyMoRS and found no significant difference (median $\Delta\alpha \approx 0.08$, $p \gg 0.05$), suggesting that particle acceleration and radiative loss processes occur in a broadly similar manner on both sides, with the observed scatter likely arising from local environmental effects rather than any intrinsic asymmetry.

\item The estimated inclination angles of the HyMoRSs in the current sample range from 22$^\circ$ to 63$^\circ$, with a mean of 38$^\circ$ and a median of 36$^\circ$. Both the mean and median values are consistent with predictions from three-dimensional relativistic hydrodynamic simulations.

\item Our analysis of the host galaxy properties of HyMoRSs indicates that they are predominantly associated with star-forming galaxies. The mid-infrared SFRs of the combined HyMoRS sample range from \(0.2\) to \(55 \, M_{\odot} \, \text{yr}^{-1}\), with a mean of \(11 \, M_{\odot} \, \text{yr}^{-1}\) and a median of \(5.45 \, M_{\odot} \, \text{yr}^{-1}\). A majority of these sources exhibit significantly high SFRs, indicative of abundant cold gas supplies fueling both star formation and AGN activity.

\item The stellar masses of HyMoRSs range from $9.35 \times 10^{8}$ to $1.85 \times 10^{11}~M_{\odot}$, with a mean value of $2.29 \times 10^{10}~M_{\odot}$ and a median of $1.17 \times 10^{10}~M_{\odot}$, consistent with the typical stellar content of massive elliptical galaxies ($10^{9}$--$10^{12}~M_{\odot}$).

\item We estimated the MIR luminosities at 3.4~$\mu$m, 4.6~$\mu$m, 12~$\mu$m, and 22~$\mu$m wavelengths. In the W1 band (3.4~$\mu$m), the luminosities range from \(4.32 \times 10^{9}\) to \(4.74 \times 10^{11}~L_\odot\), with a mean of \(6.12 \times 10^{10}~L_\odot\) and a median of \(3.98 \times 10^{10}~L_\odot\). At 4.6~$\mu$m (W2), the luminosities vary between \(2.35 \times 10^{9}\) and \(5.57 \times 10^{11}~L_\odot\), with a mean of \(4.13 \times 10^{10}~L_\odot\) and a median of \(2.12 \times 10^{10}~L_\odot\). For the W3 band (12~$\mu$m), the luminosities span from \(3.24 \times 10^{8}\) to \(6.00 \times 10^{11}~L_\odot\), with a mean of \(4.28 \times 10^{10}~L_\odot\) and a median of \(1.34 \times 10^{10}~L_\odot\). Finally, in the W4 band (22~$\mu$m), the luminosities range between \(4.36 \times 10^{8}\) and \(1.06 \times 10^{12}~L_\odot\), with a mean of \(1.04 \times 10^{11}~L_\odot\) and a median of \(4.23 \times 10^{10}~L_\odot\).

\item At least 9 HyMoRSs (25\%) in the present sample reside in dense cluster environments and are hosted by BCGs with $M_{500}$ ranging from \( 0.5 \times 10^{14} \, M_\odot \) to \( 2.27 \times 10^{14} \, M_\odot \). Only one source (J0039--0152) exceeds the \( 2 \times 10^{14} \, M_\odot \) threshold commonly used to define massive clusters. Sixteen HyMoRSs from the catalog of thirty-six sources in the present article are bent from their main axis of propagation. The bending angle for fourteen sources is between 10$^{\circ}$--20$^{\circ}$, while the remaining two sources (J0954--0003 and J2357--0725) are bent at 52$^{\circ}$ and 60$^{\circ}$.

\end{itemize}

\section*{Acknowledgements}
We thank the SARAO's team of engineers and commissioning scientists for years of intense and successful work towards delivering the wonderful MeerKAT telescope. The MeerKAT telescope is operated by the South African Radio Astronomy Observatory, which is a facility of the National Research Foundation, an agency of the Department of Science and Innovation. 
The MeerKAT data were processed using the MALS computing facility at IUCAA (\url{https://mals.iucaa.in/releases}). SK gratefully acknowledges the Department of Science \& Technology, Government of India for financial support, vide reference no. DST/WISE-PhD/PM/2023/3 (G) under the ‘WISE Fellowship for Ph.D.’ program to carry out this work.

%\bibliography{sample631}{}
%\bibliographystyle{aasjournal}

\end{document}